\documentclass[twocolumn,showpacs,prb,superscriptaddress]{revtex4}

\usepackage{graphicx}
\usepackage{color}
\usepackage{tabularx}
\usepackage{epsfig}
\usepackage{amsmath}
\usepackage{amssymb}
\usepackage{graphicx}
\usepackage{dcolumn}
\usepackage{bm}
\usepackage{wasysym}
\usepackage{pgf}

\definecolor{grn}{rgb}{0,0,0.54}

\newcommand{\ket}[1]{\ensuremath{| #1 \rangle}}

\begin{document}

\title{Mechanisms for Spin-Supersolidity in $S=1/2$
Spin-Dimer Antiferromagnets }
%%%%%%%%%%%%%%%%%%%%%%%%%%%%%%%%%%%%%%%%%%%%%%%%% 

\author{J.-D. Picon}
\affiliation{Theoretische Physik, ETH Z\"urich, 8093 Z\"urich, Switzerland} 
\affiliation{Institute of Theoretical Physics, \'Ecole polytechnique f\'ed\'erale de Lausanne, Switzerland}

\author{A. F. Albuquerque}
\affiliation{Theoretische Physik, ETH Z\"urich, 8093 Z\"urich, Switzerland} 
\affiliation{School of Physics, The University of New South Wales, Sydney, NSW 2052, Australia}

\author{K. P. Schmidt}
\affiliation{Lehrstuhl f\"ur theoretische Physik I,Otto-Hahn-Stra\ss e 4, TU Dortmund, D-44221
Dortmund, Germany}

\author{N. Laflorencie}
\affiliation{Laboratoire de Physique des Solides, Universit\'e Paris-Sud, UMR-8502 CNRS,
91405 Orsay, France}

\author{M. Troyer}
\affiliation{Theoretische Physik, ETH Z\"urich, 8093 Z\"urich, Switzerland} 

\author{F. Mila}
\affiliation{Institute of Theoretical Physics, \'Ecole polytechnique f\'ed\'erale de Lausanne, Switzerland} 

%%%%%%%%%%%%%%%%%%%%%%%%%%%%%%%%%%%%%%%%%%%%%%%%% 

\date{\today}
\pacs{03.75.Nt, 05.30.Jp, 75.10.Jm, 75.40.Mg}

\begin{abstract}
Using perturbative expansions and the Contractor Renormalization (CORE) algorithm, we
obtain effective hard-core bosonic Hamiltonians describing the low-energy physics of $S=1/2$
spin-dimer antiferromagnets known to display supersolid phases under an applied magnetic field.
The resulting effective models are investigated by means of mean-field analysis and Quantum
Monte Carlo simulations. A ``leapfrog mechanism", through means of which extra singlets delocalize in a checkerboard-solid environment via correlated hoppings, is unveiled that accounts for the supersolid behavior.
%We also deduce the correct temperature scale missed in previous studies.
\end{abstract}

\maketitle

%%%%%%%%%%%%%%%%%%%%%%%%%%%%%%%%%%%%%%%%%%%%%%%%% 

\section{Introduction}
\label{sec:intro}

Concepts and techniques developed within a well established research field
are often employed in exploring new physics displayed by apparently unrelated
systems. Following this trend, there has been an increased interest in field-induced
Bose-Einstein condensation of magnons in quantum magnets (for a recent
review, see Ref.~\onlinecite{giamarchi:07}). Although the analogy is never
complete, this line of research undoubtedly has led to considerable success
in unveiling new phenomena in a growing number of magnetic insulators under
applied magnetic field. The success of this approach suggests that one might
be able to experimentally observe more elusive bosonic behavior in quantum magnets,
such as the phase simultaneously displaying diagonal and off-diagonal order known
as {\em supersolid}.

Supersolidity has attracted enormous interest since the detection of non-classical
rotational inertia in solid Helium by Kim and Chan.\cite{kim:04a, kim:04b} Although the
correct interpretation of these measurements is still hotly debated and there seems to be
no consensus on the possibility of supersolidity in translationally invariant systems\cite{sasaki:06,bouchaud:07,pollet:07, pollet:08},
the occurrence of supersolid phases for bosonic models on a lattice is a well established
fact. While the simplest model of interacting hard-core bosons on a square lattice is
unstable against phase separation, which prevents supersolid behavior,\cite{batrouni:00,schmid:02,
sengupta:05} it has been shown that frustration\cite{wessel:05b, heidarian:05, melko:05,
boninsegni:05}, removal of the hard-core constraint\cite{sengupta:05, batrouni:06} or
inclusion of generalized couplings in the Hamiltonian\cite{chen:08, schmidt:08} can
stabilize supersolidity. Although a more direct implementation of these models remains
elusive, due to the short ranged nature of interactions between atoms in optical lattices,
one might expect that they are relevant in the context of quantum magnets under an
applied magnetic field.

Indeed, as it was first shown by Ng and Lee\cite{ng:06} and further verified by some of
us,\cite{laflorencie:07} an $S=1/2$ spin-dimer model on the square lattice with intra-plane
coupling Ising-like anisotropy [see Eq.~(\ref{eq:spin-dimer}) below] has a phase simultaneously
displaying diagonal and off-diagonal order, the equivalent of a supersolid for spin systems
(henceforth dubbed {\em spin-supersolid}). Later spin-supersolidity was also shown to occur for
$S=1$ systems on a bilayer\cite{sengupta:07a} and on a chain.\cite{sengupta:07b} However,
the exact relationship between these spin models and the aforementioned bosonic lattice
models is not well understood. For instance, one might naively expect that the $S=1/2$ spin-dimer
model investigated in Refs.~\onlinecite{ng:06} and \onlinecite{laflorencie:07} will map
onto a $t-V$ model for hard-core bosons on a square lattice, which is known not to display a
supersolid phase.\cite{schmid:02,sengupta:05} Therefore, in order to understand the mechanism
behind supersolidity in this model one should analyze the presence of extra terms in the
effective model.

Using a perturbative analysis and the contractor renormalization (CORE)
method, we derive effective Hamiltonians for the $S=1/2$ spin-dimer model studied in
Refs.~\onlinecite{ng:06} and \onlinecite{laflorencie:07}. A mean-field
analysis of the resulting generalized hard-core bosonic Hamiltonian leads
to a minimal model capable of accounting for supersolid behavior, which
is then studied by means of Quantum Monte Carlo (QMC).

%%%%%%%%%%%%%%%%%%%%%%%%%%%%%%%%%%%%%%%%%%%%%%%%% 
\begin{figure}[!tbp]
  \includegraphics*[width=0.45\textwidth]{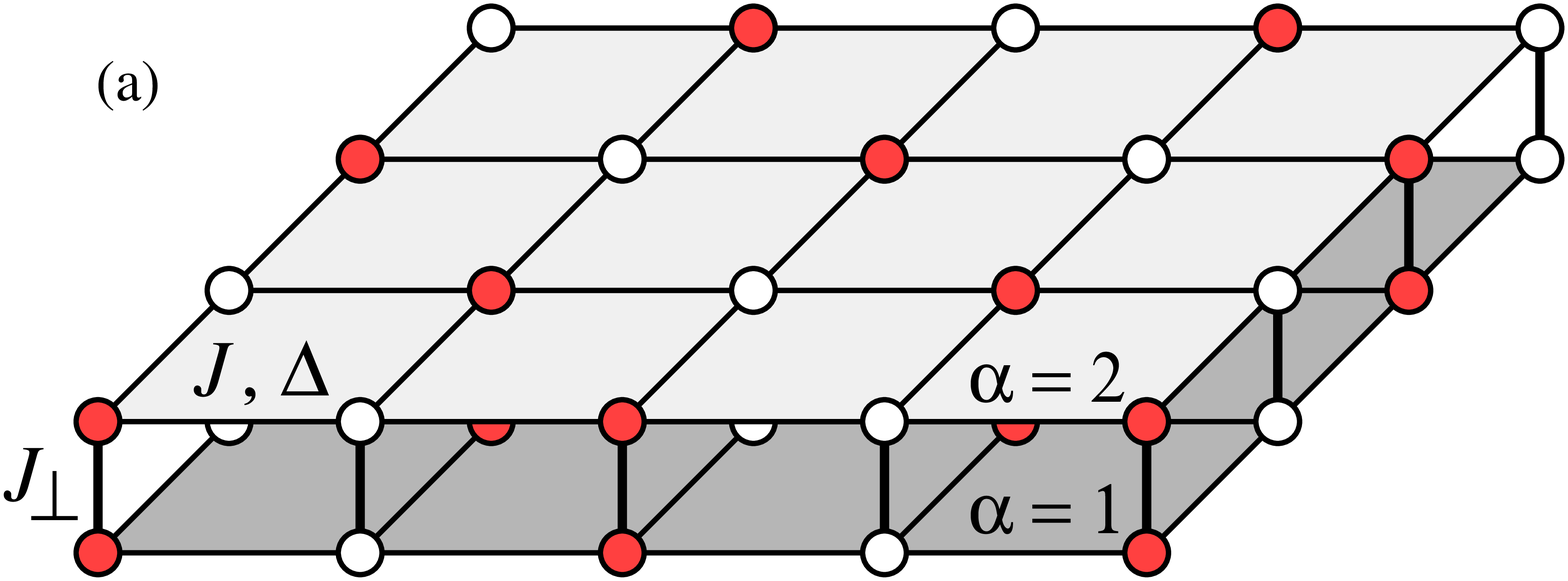}

  \vspace{0.3cm}

  \includegraphics*[width=0.45\textwidth]{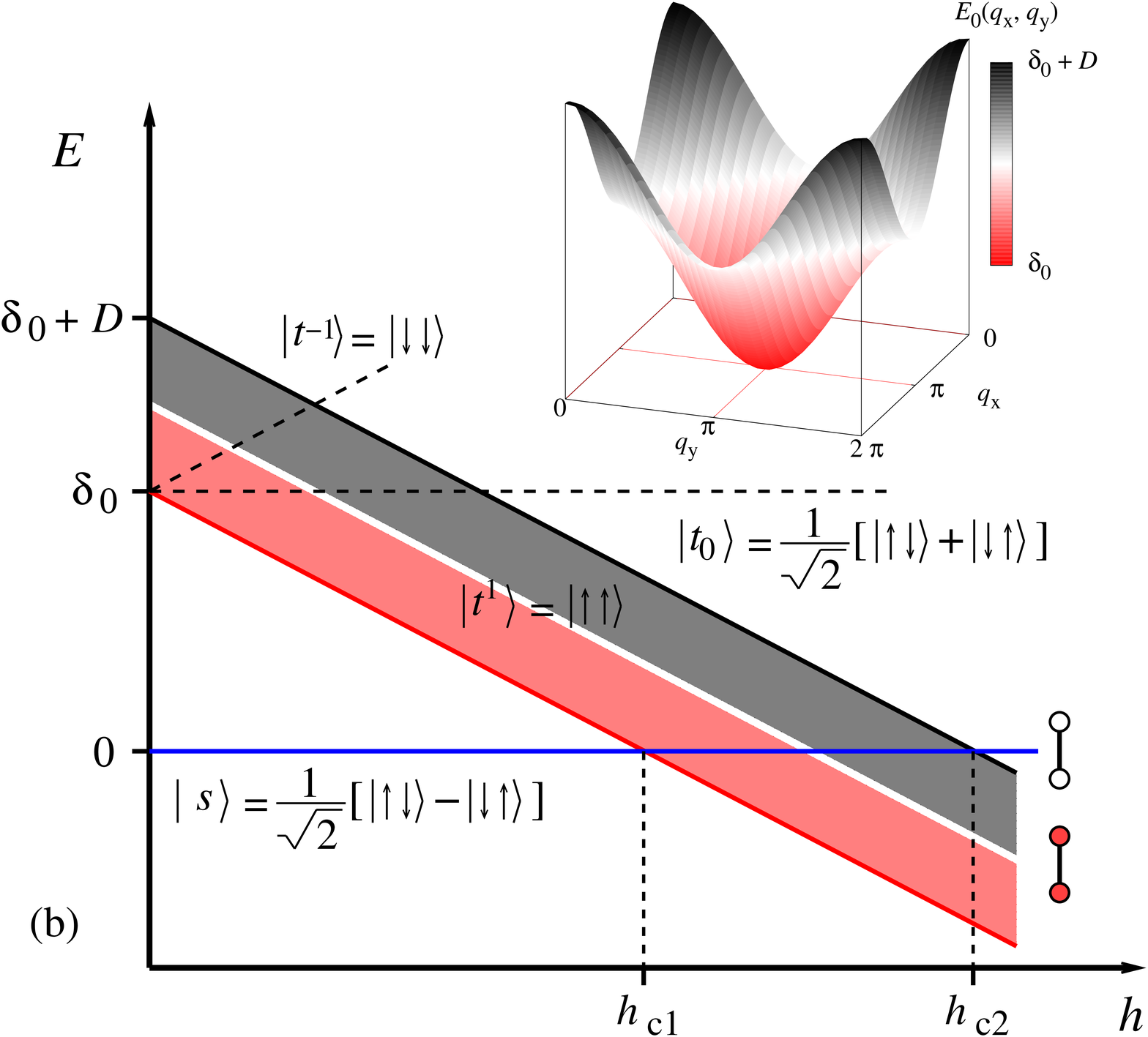}

  \caption{
    (Color online) (a) The bilayer spin system investigated in this paper, described by the
    Hamiltonian of Eq.~(\ref{eq:spin-dimer}). The strong coupling $J_{\perp}$, represented by
    thick vertical lines, accounts for the system's strong dimer character. Application of
    a magnetic field along the $z$ direction promotes dimers from a singlet
    ($\ket{s}$, vertical pairs of white circles) to a triplet ($\ket{t^{1}}$, pairs of red circles) state and controls
    the density of emergent bosons as depicted in (b). Solid (and supersolid) phases might be stabilized
    for field values in the range between the lower-critical field $h_{c1}$, where the bottom of the
    triplet band (represented in the inset and separated from the singlet state by a zero-field gap $\delta_0$ and with
    width $D$) and the singlet state become degenerate, and the upper-critical field $h_{c2}$ where the system
    becomes fully polarized. In-plane coupling $J$ leads to interactions and hopping amplitudes
    for emergent bosons, the anisotropy $\Delta$ being necessary to stabilize a checkerboard solid
    represented in the upper panel. 
  }
  \label{fig:bilayer}
\end{figure}

%%%%%%%%%%%%%%%%%%%%%%%%%%%%%%%%%%%%%%%%%%%%%%%%% 
\begin{figure}
  \begin{center}
    \includegraphics*[width=0.33\textwidth,angle=270]{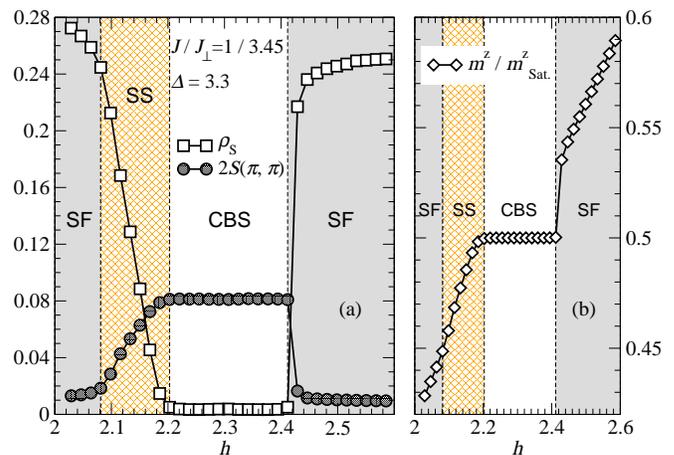}
  \end{center}
  \caption{(Color online) Quantum Monte Carlo results for the spin-dimer model Eq.~(\ref{eq:spin-dimer}).
  Simulations were performed on a $16\times 16\times 2$ lattice with $\beta=32$, for $J / J_{\perp} = 1/3.45
  \approx 0.29$ and $\Delta= 3.3$. (a) Superfluid density $\rho_{\rm S}$ (open squares) and static structure factor
  $2\times S(\pi,\pi)$ (filled circles). (b) normalized magnetization per site $m^{z}/m^{z}_{\rm sat}$ (open diamonds),
  equivalent to particle density in the bosonic language. Different phases are stabilized as a function of the
  applied magnetic field, namely: a superfluid (SF) phase with finite $\rho_{\rm S}$ and vanishing structure
  factor, an extended supersolid (SS) in which both $\rho_{\rm S}$ and $S(\pi,\pi)$ are finite and a checkerboard
  solid (CBS). Error bars are much smaller than the depicted symbols.(Adapted from Ref.~\onlinecite{laflorencie:07}.)}
  \label{fig:QMC-spin}
\end{figure}
%%%%%%%%%%%%%%%%%%%%%%%%%%%%%%%%%%%%%%%%%%%%%%%%% 

\section{The Model}
\label{sec:model}

We analyze the $S=1/2$ spin-dimer Hamiltonian analyzed by Ng and Lee\cite{ng:06} and some
of us,\cite{laflorencie:07} which reads
\begin{equation}
  \begin{split}
    {\mathcal H} =  J_{\perp} \sum_{i} \vec{S}_{i,1} \cdot \vec{S}_{i,2}
    - h \sum_{i, \alpha=1,2} S^{z}_{i, \alpha}\\
    +J_{} \sum_{\left\langle i,j \right\rangle, \alpha=1,2}
    (S^{x}_{i,\alpha}S^{x}_{j,\alpha} + S^{y}_{i,\alpha}S^{y}_{j,\alpha}
    +\Delta S^{z}_{i,\alpha}S^{z}_{j,\alpha})~.
  \end{split}
  \label{eq:spin-dimer}
\end{equation}
$\vec{S}_{i,\alpha}$ is an $S=1/2$ operator attached to the site $i$ of the layer $\alpha$ (see Fig.~\ref{fig:bilayer}).
$J_{\perp}$ couples spins in different layers and is considered to be the essential coupling, being responsible
for the system's strong dimerized character (we set $J_{\perp} = 1$ throughout the rest of the
paper). Spins in the same layer interact via the coupling $J$ and $\Delta$ is an Ising-like
anisotropy; finally, the magnetic field $h$ is applied along the easy-axis. We will mainly focus on
the set of parameters considered in Refs.~\onlinecite{ng:06} and \onlinecite{laflorencie:07},
$J / J_{\perp} = 0.29$ and $\Delta= 3.3$, leading to an extended supersolid phase as evident from
the QMC results for the spin stiffness $\rho_{\rm S}$ and static structure factor
$S(\pi,\pi)$ obtained by Laflorencie and Mila\cite{laflorencie:07} and reproduced in 
Fig.~\ref{fig:QMC-spin}.

\begin{figure}
  \begin{center}
    \includegraphics*[width=0.4\textwidth]{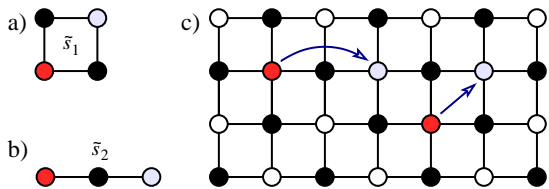}
  \end{center}
  \caption{(Color online) (a) NNN correlated hopping with amplitude $\tilde{s}_{1}$ and (b) third-neighbor
  correlated hopping with amplitude $\tilde{s}_{2}$: singlets (holes) hop in between red and light-blue sites
  {\em only} if the black-filled circles are occupied by holes [in (a), at least one of the sites must be occupied;
  if both are, the process occurs with amplitude $2\tilde{s}_{1}$]. This ``leapfrog mechanism" allows for
  extra holes to delocalize in a checkerboard ordered environment, as illustrated in (c), and to condense,
  giving rise to supersolid behavior.}
  \label{fig:leapfrog}
\end{figure}

Our goal is to show that we can understand the emergence of SS for the spin model
Eq.~(\ref{eq:spin-dimer}) in terms of simple microscopic mechanisms. In achieving this, we
derive effective bosonic models for Eq.~(\ref{eq:spin-dimer}) by means of two different
procedures: high-order perturbative series expansions and the Contractor Renormalization
algorithm (CORE). We are going to show that correlated hoppings for singlets (holes) with amplitudes
$\tilde{s}_{1}$ (next-nearest-neighbor, NNN, hopping which occurs only if at least one of the other sites on the same
plaquette is occupied) and $\tilde{s}_{2}$ (assisted third-neighbor hopping occurring only when
the site in between is occupied), depicted in Fig.~\ref{fig:leapfrog} (a) and (b) respectively,
are crucial in accounting for SS behavior for the model of Eq.~(\ref{eq:spin-dimer}). It is easy to see
[Fig.~\ref{fig:leapfrog}(c)] that these processes prevent phase separation\cite{sengupta:05}
in the hard-core bosonic model on the square lattice ($t-V$ model) by allowing extra singlets (holes) to
delocalize in a checkerboard solid (CBS) environment by ``leapfrogging" on the other
sublattice and forming a condensate. It is useful to define the quantity we call ``leapfrog
ratio"
\begin{equation}
  \Sigma = \frac{(2|\tilde{s}_{1}| + |\tilde{s}_{2}|)} {|\tilde{t}_{1}|}~,
  \label{eq:leapfrog_ratio}
\end{equation}
where $\tilde{t}_{1}$ is the nearest-neighbor (NN) hopping amplitude for holes. It was
shown by Sengupta {\em et al.}\cite{sengupta:05} that the energetic gain in the domain wall formation
behind phase separation in the $t-V$ model is $c\tilde{t}_1$, where $c$ lies in the interval $[1,2]$.
Therefore, for a system of hard-core bosons on the square lattice, the energetic
gain associated to the correlated hoppings depicted in Fig.~\ref{fig:leapfrog}
must be larger than $c\tilde{t}_{1}$, implying that the condition $\Sigma > c/4$ must be obeyed,
for SS behavior to emerge.

\section{Pertubative Expansions}
\label{sec:perturbation}

Following the work of Totsuka\cite{totsuka:98} and Mila,\cite{Mila:1998} we restrict ourselves to the limit where $J_\perp$ is the
main energy scale and the system consists of weakly coupled dimers. The application of a magnetic field
lowers the energy of one of the triplet bands and at the critical field $h_{\rm c1}$ the 
singlet state $\ket{s}$ (holes) and the bottom of the triplet $\ket{t^1}$ (bosons) band become
degenerate [see Fig.~\ref {fig:bilayer}(b)]. By expanding the Hamiltonian Eq.~(\ref{eq:spin-dimer})
in terms of the small parameter $J/J_\perp$ we can thus obtain an effective hard-core bosonic model.
Within first-order in $J/J_\perp$, the only effective couplings in the model obtained in this way are nearest-neighbor
hopping amplitude $t_{1}$ and repulsion $V_{1}$ for the emergent bosons (triplets).\cite{Mila:1998} However,
since this so-called $t-V$ model [equivalent to Eq.~(\ref{eq:SE-boson})
below if we set $s_{1,2}^{\rm P2} = 0$] is known to display {\it no} SS phase,\cite{sengupta:05} higher-order
effective couplings should be taken into account and we proceed to their derivation.

%%%%%%%%%%%%%%%%%%%%%%%%%%%%%%%%%%%%%%%%%%%%%%%%% 
%%%%%%%%%%%%%%%%%%%%%%%%%%%%%%%%%%%%%%%%%%%%%%%%%

\subsection{Second-Order Expansion}
\label{sec:2nd}

We extend the perturbative analysis of Mila\cite{Mila:1998} to second order in $J/J_\perp$,
obtaining the following effective Hamiltonian 
\begin{equation}
  \begin{split}
        \mathcal{H}^{\rm P2}_{\rm eff} = -\mu^{\rm P2} \sum_{i} n_{i} 
        + \sum_{\left\langle i,j \right\rangle} \left[ t_1^{\rm P2} (b_{i}^\dagger b_{j} + {\rm H.c.})+V_1^{\rm P2} n_{i}n_{j} \right] \\
        + s_{1,2}^{\rm P2} \sum_{\left\langle i,j,k \right\rangle} \left[ b_{i}^\dagger (1-n_{j})b_{k} +{\rm H.c.} \right] ~,
  \end{split}
  \label{eq:SE-boson}
\end{equation}
with effective couplings (we set $J_\perp=1$)
\begin{equation}
  \begin{split}
  \mu^{\rm P2} =1+ \dfrac{ {J}^2(2+\Delta^2)}{4}+h~,\\
  t_1^{\rm P2} = J/2,~~~s_{1,2}^{\rm P2}=-{J}^2/16~, \\
  V_1^{\rm P2}= \dfrac{J\Delta}{2}-\dfrac{{J}^2(2+\Delta^2)}{8}~.
  \end{split}
  \label{eq:SE_couplings}
\end{equation}
In Eq.~(\ref{eq:SE-boson}), $n_{i} = b_{i}^\dagger b_{i}=\{0,1\}$ is the occupation number for hardcore bosons ($\ket{t^1}$ triplets) at
the site $i$ on the square lattice formed by the spin-dimers. $\left\langle i,j \right\rangle$ denotes
nearest-neighbor (NN) sites on this lattice and $\left\langle i,j,k \right\rangle$ is such that $j$ is a
common nearest-neighbor for the second- or third-neighbor sites $i$ and $k$.\cite{second-order}
The physical processes at play become more evident after applying a particle-hole transformation,
$(1-n_i) \rightarrow \tilde{n}_i $ and $b^\dagger \rightarrow \tilde{b}$, to Eq.~(\ref{eq:SE-boson}):
\begin{equation}
  \begin{split}
        \tilde{\mathcal{H}}^{\rm P2}_{\rm eff} =\tilde{\mu}^{\rm P2} \sum_{i} \tilde{n}_{i} 
        + \sum_{\left\langle i,j \right\rangle} \left[ t_1^{\rm P2} (\tilde{b}_{i}^\dagger \tilde{b}_{j} + {\rm H.c.})+V_1^{\rm P2}
        \tilde{n}_{i}\tilde{n}_{j} \right] \\
        + s_{1,2}^{\rm P2} \sum_{\left\langle i,j,k \right\rangle} \left( \tilde{b}_{i}^\dagger\tilde{n}_{j}\tilde{b}_{k} +{\rm H.c.} \right) ~.
  \end{split}
  \label{eq:SE-holes}
\end{equation}
We have ignored constant terms and $\tilde{\mu}^{\rm P2} = \mu^{\rm P2}+2V_1^{\rm P2}$;
$\tilde{n}_i = \tilde{b}_{i}^\dagger \tilde{b}_{i}$ is now the {\em singlet} (holes) occupation
number. In addition to the
first-order couplings $t_1^{\rm P2}$ (NN hopping amplitude) and $V_1^{\rm P2}$ (NN repulsion)
the second-order effective Hamiltonian also contains a correlated hopping term with amplitude $s_{1,2}^{\rm P2}$
[see Fig.~\ref{fig:leapfrog}(a-b)].\cite{second-order} Correlated hoppings have been shown to stabilize
supersolidity\cite{schmidt:08} by allowing particles to delocalize in a CBS ordered background. However, the second-order
amplitude for correlated hoppings is too small\cite{sengupta:05,schmidt:08} to prevent phase separation:
the ``leapfrog ratio" of Eq.~(\ref{eq:leapfrog_ratio}), $\Sigma^{\rm P2} = 3J/8 \approx 0.12$ for
$J / J_{\perp} = 0.29$ is too small and  
%, is much smaller than $c/4$ ($c \in [1,2]$) and 
cannot account for supersolidity. Therefore we extend our analysis and include higher-order corrections to
the parameters $t_1^{\rm P2}$, $V_1^{\rm P2}$ and $s_{1,2}^{\rm P2}$
in Eq.~(\ref{eq:SE-holes}) with the help of perturbative continuous unitary transformations (PCUTs).

%%%%%%%%%%%%%%%%%%%%%%%%%%%%%%%%%%%%%%%%%%%%%%%%% 
\begin{figure}[!tbp]
  \includegraphics*[width=0.33\textwidth,angle=270]{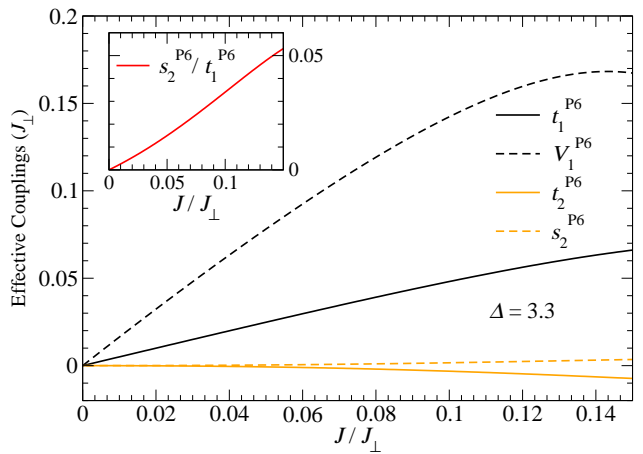}

  \caption{(Color online) $\Delta=3.3$. Effective couplings obtained from the PCUTs procedure described in
  the main text: nearest-neighbor hopping amplitude $t_1^{\rm P6}$ (solid dark line) and interaction $V_1^{\rm P6}$
  (dashed dark line) and amplitudes for next-nearest-neighbor uncorrelated ($t_2^{\rm P6}$, solid light line)
  and correlated ($s_2^{\rm P6}$, dashed light line) hoppings, as a function of $J/J_{\perp}$. The inset
  shows the dependence of the ratio $s_2^{\rm P6} / t_1^{\rm P6}$ on $J/J_{\perp}$.
  }
  \label{fig:PCUT}
\end{figure}
%%%%%%%%%%%%%%%%%%%%%%%%%%%%%%%%%%%%%%%%%%%%%%%%%

\subsection{Perturbative Continuous Unitary Transformations (PCUTs)}
\label{sec:PCUTs}

The method of continuous unitary transformations (CUTs)\cite{wegner:94,glazek:93,glazek:94,knetter:01b}
in its perturbative variant\cite{stein:97,mielke:98,uhrig:98c,knetter:00a} and quasi-particle conserving form
is an efficient tool to derive effective low-energy models for coupled quantum dimer networks in a magnetic
field up to high order in perturbation.\cite{mila:08,dorier:08} 

To this end, the original spin Hamiltonian Eq.~(\ref{eq:spin-dimer}) is rewritten in terms of rung triplet
operators $t^{(\dagger)}_\alpha$ with $\alpha=\{ \pm1, 0\}$. This Hamiltonian does not
conserve the number of triplets $Q=\sum_{i,\alpha=\pm 1, 0} t^\dagger_\alpha  t^{\phantom{\dagger }}_\alpha$
in the system. The basic idea of quasi-particle conserving CUTs is to transform $\mathcal{H}$ into
$\mathcal{H}_{\rm eff}$ such that $[\mathcal{H}_{\rm eff},Q]=0$, i.e. the number of quasi-particles (triplons, in
the present case), is a conserved quantity. Since the total $S^z_{\rm tot}$ is also a conserved quantity,
the magnetic field term does {\it not} change under the unitary transformation. For the case of coupled
dimers in a magnetic field, one can therefore restrict to terms in $\mathcal{H}_{\rm eff}$ consisting solely
of triplet operators $t^{(\dagger)}_1$ in order to describe the low-energy physics.

In general, a continuous parameter $l$ is introduced such that $l=0$ refers to the initially given system
$\mathcal{H}$ and $l=\infty$ corresponds to the final effective system $\mathcal{H}_{\rm eff}$. Let $U$
be the unitary transformation which diagonalizes the Hamiltonian $\mathcal{H}$ and
$\mathcal{H}(l)=U^\dagger (l)\mathcal{H}U(l)$. Then this unitary transformation is equivalent to performing
an infinite sequence of unitary transforms $e^{-\eta (l)dl}$ with the anti-hermitian generator
\begin{equation}
  \eta (l)=-U^\dagger (l)\partial_l U(l) \quad~.
\end{equation}
The derivation with respect to $l$ results in the so-called flow equation
\begin{equation}
  \partial_l \mathcal{H} (l) = [\eta (l),\mathcal{H}(l)]~,
  \label{eq:flow}
\end{equation}
which defines the change of the Hamiltonian during the flow. The properties of the effective Hamiltonian
depend strongly on the choice of the generator $\eta$. Quasi-particle-conserving CUTs chooses $\eta$
such that the Hamiltonian $\mathcal{H}_0$ maps onto an effective Hamiltonian which \emph{conserves}
the number of quasi-particles.\cite{stein:97,mielke:98,uhrig:98c,knetter:00a} 

In the following we consider the limit of weakly coupled rung dimers, i.e. we set $J_\perp=1$ and treat
$J$ and $\Delta$ as a small expansion parameters. Using a series expansion ansatz for $\eta$
and $\mathcal{H}$ in Eq.~\ref{eq:flow}, one can derive the effective quasi-particle conserving Hamiltonian
up to high order in perturbation.\cite{stein:97,knetter:00a,knetter:03a} The results are obtained in the thermodynamic
limit and in second quantization. 

We stress again that the total $S^z_{\rm tot}$ is a conserved quantity. The magnetic field term has {\it not}
changed under the unitary transformation and the low-energy physics is solely influenced by the local
singlet $|s\rangle$ and the triplet $|t^1\rangle$ polarized parallel to the magnetic field (as discussed before).
Identifying $|s\rangle$ with an empty site and $|t^1\rangle$ with the presence of a hardcore boson (as before),
we can deduce the effective Hamiltonian in this language by calculating matrix elements on finite
clusters.\cite{knetter:03a} 

We have extended the derivation of the effective parameters appearing in Eq.~(\ref{eq:SE-holes}),
now relabeled as $t_1^{\rm P6}$, $V_1^{\rm P6}$ and  $s_{2}^{\rm P6}$, to sixth order in $J / J_{\perp}$
and have additionally calculated the amplitude $t_2^{\rm P6}$ for uncorrelated third-neighbor hopping.
Explicit formulas are given in Appendix \ref{sec:appendix-PCUT} and dependences on $J / J_{\perp}$
are shown in Fig.~\ref{fig:PCUT} for $\Delta= 3.3$. Comparison between $t_1^{\rm P6}$ and results
obtained from CORE (see Fig.~\ref{fig:comparison} and discussion below) suggests that our perturbative
analysis remains valid up to $J / J_{\perp} \lesssim 0.15$.

By applying a particle-hole transformation it is possible to show that the condition $t_2^{\rm P6} = -2 s_2^{\rm P6}$,
approximately fulfilled by our results (Fig.~\ref{fig:PCUT}), implies a vanishing amplitude for {\em uncorrelated}
third-neighbor hopping for {\em holes} (singlets) and therefore the magnitude of $s_2^{\rm P6}$ is the relevant kinetic
scale for supersolidity (a similar situation happens for the effective Hamiltonian derived from CORE, see Sec.~\ref{sec:minimal}).
As we mentioned before, supersolid behavior is expected to occur for large enough values of  $s_2^{\rm P6} / t_1^{\rm P6}$.
\cite{sengupta:05,schmidt:08} However, our results for this ratio, shown in the inset of Fig.~\ref{fig:PCUT}, are
clearly too small for preventing domain wall formation,\cite{sengupta:05,schmidt:08} and therefore one does
not expect to reproduce the  extended SS phase observed for the original spin model, Eq.~(\ref{eq:spin-dimer}).
Consequently, either our idea that the model can be described by only taking into account $\ket{s}$ and $\ket{t^1}$ is wrong,
or we must go beyond a perturbative analysis. Since according to Ng and Lee \cite{ng:06} contributions from the other two triplets
states $\ket{t^0}$ and $\ket{t^{-1}}$, if non-zero, are negligible close to half-filling, we therefore resort on a non-perturbative
approach to our problem, namely the CORE algorithm. 

\section{Contractor Renormalization}
\label{sec:core}

The contractor renormalization (CORE) method was introduced by Morningstar and Weinstein
\cite{morningstar:94, morningstar:96} and has been recently\cite{abendschein:07} applied to the study
of the spin-dimer Hamiltonian described by Eq.~(\ref{eq:spin-dimer}). We extend these results
by considering the next range in the effective couplings and analyzing in more detail the resulting
effective bosonic model.

\subsection{Procedure}
\label{sec:CORE_procedure}

The basic idea behind CORE (for comprehensive accounts the reader
is referred to Refs.~\onlinecite{altman:02} and \onlinecite{capponi:06}) is to
{\em project out} high energy degrees of freedom and to derive an effective
Hamiltonian describing the low-energy physics of the original model. Usually
this is done by first decomposing the lattice on which the original model is defined
into elementary blocks and diagonalizing the Hamiltonian on a single block (while an extended
method without this restriction was introduced by some of us recently,
\cite{albuquerque:08b} here the standard CORE method is more appropriate).
After choosing a suitable number of low-energy block states, the model
is subsequently diagonalized on a cluster consisting of a few elementary
blocks and the lowest energy cluster states are projected onto the restricted
basis formed by the tensor products of the retained block states. An effective
Hamiltonian is then obtained by imposing the constraint that the low-energy
spectrum of the full problem is exactly reproduced and by subtracting shorter-range
contributions obtained from previous steps involving lesser blocks. The validity of the
procedure can be checked by either analyzing the magnitude of long-range effective
couplings (large values associated to these signal the inadequacy of
the chosen restricted set of degrees of freedom in accounting for the system's
low-energy behavior) or, perhaps more accurately, by keeping track of
the weight of the reduced density-matrix associated to a single block.
\cite{capponi:04, abendschein:07}

For the spin-dimer model considered here, Eq.~(\ref{eq:spin-dimer}),
large values for the inter-plane coupling $J_{\perp}$ imply that the
natural choice when applying CORE is to consider the dimers
as the elementary blocks: dimer singlet states, $\ket{s}$, corresponding to an unoccupied
site in the effective model living on the square lattice, and an emergent boson
created by promoting one singlet to an $S^z = 1$ triplet state, $\ket{t^{1}}$, are the retained
block states. The adequacy of this reduced set of degrees of freedom in describing the
Hamiltonian of Eq.~(\ref{eq:spin-dimer}) in the regime known\cite{ng:06, laflorencie:07} to display
supersolid behavior was verified by Abendschein and Capponi\cite{abendschein:07}
and is confirmed in the present work.

%%%%%%%%%%%%%%%%%%%%%%%%%%%%%%%%%%%%%%%%%%%%%%%%% 
\begin{figure}[!tbp]
  \includegraphics*[width=0.45\textwidth]{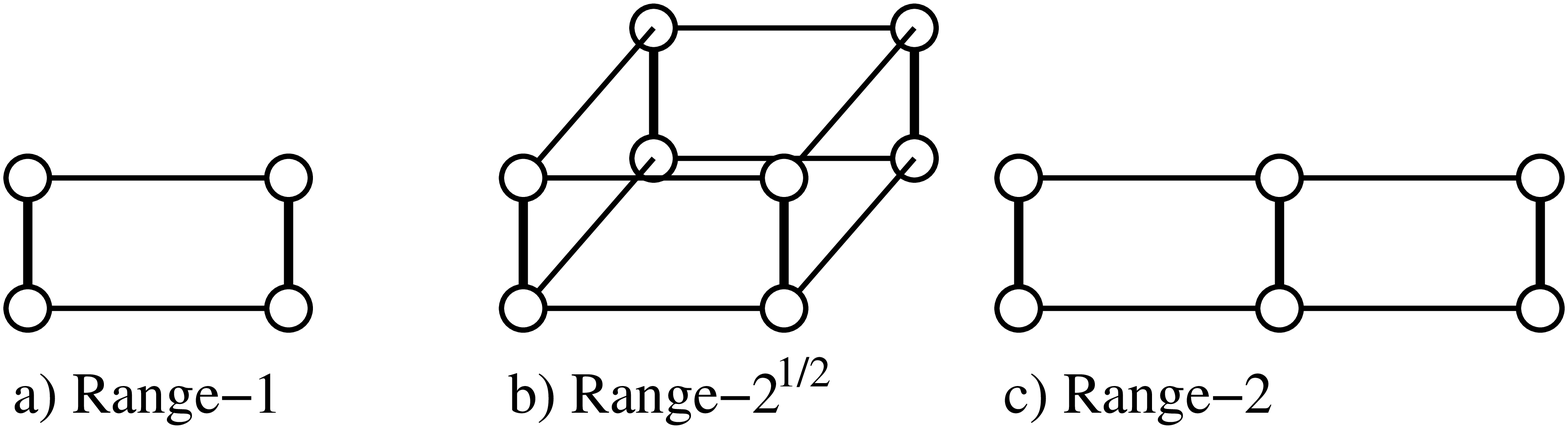}

  \caption{
    Clusters used in the CORE derivation of the effective model, labeled according
    to the longest range effective couplings on the square lattice. In this
    convention, range-1 interactions are obtained from the analysis of the cluster consisting of two
    dimers (a), range-$2^{1/2}$ from the cluster with four dimers forming a square plaquette (b) and range-2
    by considering three dimers along a line (c).
  }
  \label{fig:CORE_clusters}
\end{figure}
%%%%%%%%%%%%%%%%%%%%%%%%%%%%%%%%%%%%%%%%%%%%%%%%%

Our results are obtained from the analysis of the clusters depicted in Fig.~\ref{fig:CORE_clusters}.
They are labelled according to the maximum range for the effective couplings:
{\em range-$1$} are the results obtained from the analysis of the cluster containing two
dimers shown in Fig.~\ref{fig:CORE_clusters}(a), {\em range-$2^{1/2}$} denote the ones
from the cluster with four dimers arranged as a plaquette [Fig.~\ref{fig:CORE_clusters}(b)]
and {\em range-$2$} results from the three-dimer cluster shown in Fig.~\ref{fig:CORE_clusters}(c).
We gauge the validity of the mapping onto a system of hard-core bosons by analyzing
corrections to the nearest-neighbor (NN) hopping amplitude $t_1$ (for {\em particles}) obtained from
range-$2^{1/2}$ and range-2 CORE calculations: whenever the sum of these contributions exceeds
the value obtained from range-1 CORE we {\em assume} that a valid mapping is not obtained.
While the criteria used by Abendschein and Capponi\cite{abendschein:07} is probably
more accurate, our results agree qualitatively with theirs and suffice for our analysis. More importantly,
for the parameters ($\Delta = 3.3$ and $J/J_{\perp}=0.29$) leading to supersolidity previously considered
in the literature\cite{ng:06, laflorencie:07} both criteria validate the mapping onto the effective
bosonic model.

The effective hard-core bosonic Hamiltonian obtained from the CORE calculation is, after applying
a particle-hole transformation $(1-n_i) \rightarrow \tilde{n}_i $ and $b^\dagger \rightarrow \tilde{b}$,
given by
\begin{equation}
  \begin{split}
    \tilde{{\mathcal H}}_{\rm eff}^{\rm C} =  \sum_{i} \left\{ -\tilde{\mu}^{\rm C} \tilde{n}_{i}  +
    \left[ \tilde{{\mathcal V}}_{i} + \tilde{{\mathcal W}}_{i}\right]
    + \left[\tilde{{\mathcal T}}_{i} + \tilde{{\mathcal S}}_{i} + \tilde{{\mathcal R}}_{i} \right] \right\} ~,
  \end{split}
  \label{eq:eff_CORE}
\end{equation}
where $\tilde{\mu}^{\rm C}$ is the chemical potential for the {\em holes} (singlets). $\tilde{{\mathcal V}}$ comprises
two-body interactions and $\tilde{{\mathcal W}}$ three- and four-body interactions; $\tilde{{\mathcal T}}$, 
$\tilde{{\mathcal S}}$ and  $\tilde{{\mathcal R}}$ are the kinetic contributions: direct and correlated hopping terms.
Full expressions for each of these terms are given in Appendix~\ref{sec:appendix-CORE}.

%%%%%%%%%%%%%%%%%%%%%%%%%%%%%%%%%%%%%%%%%%%%%%%%% 
\begin{figure}[!tbp]
  \includegraphics*[width=0.33\textwidth,angle=270]{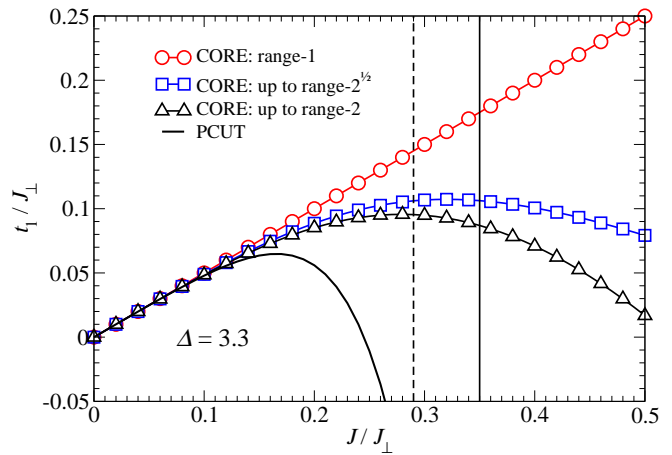}

  \caption{
    (Color online) Comparison between CORE (range-1, -$2^{1/2}$ and -2) and PCUTs results
    for the nearest-neighbor hopping amplitude $t_1$ (for {\em particles}) in the effective bosonic model as a function of
    $J/J_\perp$ for $\Delta = 3.3$, as in Refs.~\onlinecite{ng:06} and \onlinecite{laflorencie:07}. The
    value $J/J_\perp = 0.29$ is highlighted by the vertical dashed line. The vertical solid line
    indicates the point where longer range ($2^{1/2}$ and 2) corrections to $t_1$ become larger than the
    range-1 contribution, signaling the breakdown of the mapping onto a bosonic model (see main text).
  }
  \label{fig:comparison}
\end{figure}
%%%%%%%%%%%%%%%%%%%%%%%%%%%%%%%%%%%%%%%%%%%%%%%%% 

\subsection{Comparison with Pertubative Expansion}

Figure \ref{fig:comparison} shows our results for the effective nearest-neighbor hopping amplitude $t_1$
(for {\em particles}) obtained from CORE (range-1, -$2^{1/2}$ and -2) and from PCUTs for $\Delta=3.3$ and
as a function of $J/J_{\perp}$. As expected, the various ranges CORE results agree with the ones
obtained from PCUTs in the limit of small $J / J_{\perp}$, where both results are essentially exact. However,
for $J / J_{\perp} \gtrsim 0.15$ higher-order terms in the perturbative expansion start to dominate,
invalidating the PCUTs analysis. Crucially, for the value $J / J_{\perp} = 0.29$ considered in
Refs.~\onlinecite{ng:06} and \onlinecite{laflorencie:07} (highlighted by the vertical dashed line
Fig.~\ref{fig:comparison}), the PCUTs expansion is clearly invalid, while longer-range CORE results are essentially converged.

These results can be understood if we remark that any perturbative expansion about the weakly coupled dimer limit is
only valid as long as one stays in the zero-field rung-singlet phase, with a finite gap to all three triplet modes.
However, it has been shown\cite{ng:06, laflorencie:07} that for $\Delta= 3.3$ and $J/J_\perp = 0.29$ the
zero-field ground-state of the spin-dimer model Eq.~(\ref{eq:spin-dimer}) displays long-range N\'{e}el order implying the existence of a quantum critical point $J_{\rm c} (h=0)/J_\perp <0.29$
(evident from poles in Pad\'{e} analysis for the perturbation series) beyond which our perturbative expansions become meaningless.
On the other hand, although CORE relies on a strong dimerized character (so that dimer singlets and triplets are
the relevant local degrees of freedom), it does not assume any particular ordering and therefore remains valid
across the critical regime.

%%%%%%%%%%%%%%%%%%%%%%%%%%%%%%%%%%%%%%%%%%%%%%%%% 
\begin{table}
  \begin{tabular}{|c|c|c|c|}\hline
    \multicolumn{4}{|c|} {$\tilde{\mu}^{\rm C} = 2.911009-h$}   \\\hline
    $\tilde{V}_1^{\rm C}$ & 0.336874 & $\tilde{V}_2^{\rm C}$ & -0.008851 \\
    $\tilde{V}_3^{\rm C}$ & -0.011122 & $\tilde{W}_1^{\rm C}$ & 0.009035 \\
    $\tilde{W}_2^{\rm C}$ & -0.002257   & $\tilde{W}_3^{\rm C}$ & -0.064354 \\\hline
    $\tilde{t}_1^{\rm C}$ & 0.145 & $\tilde{t}_2^{\rm C}$ & 0 \\
    $\tilde{s}_1^{\rm C}$ & -0.017190 & $\tilde{s}_2^{\rm C}$ & -0.021471 \\
    $\tilde{s}_3^{\rm C}$ & -0.009378 & $\tilde{s}_4^{\rm C}$ & -0.000678 \\
    $\tilde{s}_5^{\rm C}$ & -0.005367 & $\tilde{s}_6^{\rm C}$ & -0.009977 \\
    $\tilde{r}_1^{\rm C}$ &0.000988 & $\tilde{r}_2^{\rm C}$ & -0.008850 \\\hline
  \end{tabular}
  \caption{Couplings in the effective Hamiltonian obtained from CORE [up to range-2, Eqs.~(\ref{eq:eff_CORE},
  \ref{eq:two-body}-\ref{eq:double_hoppings})] for $\Delta = 3.3$ and $J/J_{\perp}=0.29$.
  Units are set by $J_{\perp}=1$.}
  \label{tab:revsymb}
\end{table}
%%%%%%%%%%%%%%%%%%%%%%%%%%%%%%%%%%%%%%%%%%%%%%%%% 
%%%%%%%%%%%%%%%%%%%%%%%%%%%%%%%%%%%%%%%%%%%%%%%%% 
\begin{figure}
  \begin{center}
    \includegraphics*[angle=270,width=0.45\textwidth]{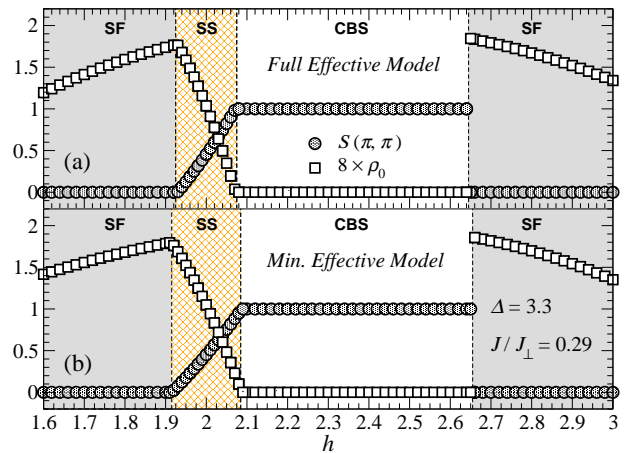}
  \end{center}
 \caption{(Color online) $\Delta=3.3$, $J/J_\perp=0.29$. Mean-field results for the CBS structure factor $S(\pi,\pi)$
 (filled circles) and condensate density $\rho_0$ (open squares) as a function of the magnetic field $h$
 for: (a) the full range-2 CORE effective Hamiltonian, Eqs.~(\ref{eq:eff_CORE}, \ref{eq:two-body} - \ref{eq:double_hoppings}),
 and (b) the minimal model of Eq.~(\ref{eq:CORE-minimal-hole}), with NN hopping amplitude given by
 Eq.~(\ref{eq:t1_min}). Values for the effective couplings are shown in Table \ref{tab:revsymb} and
 successive phases are labeled as: condensate, supersolid (${\rm SS}$) and checkerboard solid (${\rm CBS}$).}
  \label{fig:MF-CORE}
\end{figure}
%%%%%%%%%%%%%%%%%%%%%%%%%%%%%%%%%%%%%%%%%%%%%%%%% 

\section{Mechanism for Spin-Supersolidity}
\label{sec:analysis}

Numerical values obtained from CORE for {\em all} effective couplings (up to range-2) appearing in
Eqs.~(\ref{eq:eff_CORE}, \ref{eq:two-body}-\ref{eq:double_hoppings}) are shown in Table \ref{tab:revsymb}
for the parameters $\Delta=3.3$, $J/J_\perp=0.29$ used in the original QMC simulations.\cite{ng:06,laflorencie:07}
We use the mean-field (MF) approach discussed in Appendix \ref{sec:appendix-MF} and calculate
the dependence of the condensate density $\rho_0$ and CBS order parameter [see Eqs.~(\ref{eq:MF_rhoS},
\ref{eq:MF_CBS})] on magnetic field $h$. The results are shown in Fig.~\ref{fig:MF-CORE}(a).
The semi-quantitative agreement between these results and the QMC data for the original model
Eq.~(\ref{eq:spin-dimer}) (shown in Fig.~\ref{fig:QMC-spin}) is remarkable if we keep in mind
that only contributions of up to range-2 have been considered in the CORE calculation. However,
MF approaches are known to overestimate supersolid behavior\cite{schmid:02,sengupta:05, schmidt:08} and
the effects of quantum fluctuations must be carefully analyzed.

Unfortunately, the effective Hamiltonian obtained from CORE, Eqs.~(\ref{eq:eff_CORE},
\ref{eq:two-body}-\ref{eq:double_hoppings}), is complex and poses great challenges for more unbiased analysis.
We therefore use the aforementioned MF procedure in gauging the relative importance of each term, with a twofold
purpose: (a) identifying the dominant mechanism accounting for supersolidity in the spin-dimer model of
Eq.~(\ref{eq:spin-dimer}) and (b) obtaining a simpler effective model amenable to QMC simulations (see below)
in order to check whether the conjectured mechanisms survive after quantum fluctuations are taken into account.

\subsection{Minimal Hamiltonian}
\label{sec:minimal}

In deciding on a minimal model we should obviously take into account the magnitudes associated with
each term in Eqs.~(\ref{eq:eff_CORE}, \ref{eq:two-body}-\ref{eq:double_hoppings}): we start by neglecting
{\em all} effective couplings smaller than $0.1 \tilde{t}_1^{\rm C}$, where $\tilde{t}_1^{\rm C}$ is the NN hopping
for holes (singlets). Furthermore, since SS takes place only close to half-filling, we can also neglect the four-body
term with coupling $\tilde{W}_3^{\rm C}$ [see Eq.(\ref{eq:multi-body})]. The resulting model is identical to the
second-order effective Hamiltonian [Eq.~(\ref{eq:SE-holes})],\cite{second-order} but with strongly renormalized couplings.
In particular, the couplings associated to the correlated hoppings $\tilde{s}_{1}^{\rm C}$ and $\tilde{s}_{2}^{\rm C}$
[see Fig.~\ref{fig:leapfrog} (a,b)] are considerably larger than predicted by the perturbative analysis,
\cite{second-order} as required for SS to emerge. However, the MF analysis of the resulting model
shows that the extra kinetic energy associated to the large effective amplitudes for correlated
hoppings requires the addition of the {\em attractive} two-body interactions $\tilde{V}_2^{\rm C}$ and
$\tilde{V}_3^{\rm C}$ (see Table \ref{tab:revsymb}) to stabilize a CBS plateau. These considerations lead
to the {\em minimal
model}:
\begin{equation}
  \begin{split}
        \tilde{\mathcal{H}}^{\rm C}_{\rm min} = -\tilde{\mu}^{\rm C} \sum_{i} \tilde{n_{i}} 
        + \sum_{\left\langle i,j \right\rangle} \left[ \tilde{t}_1^{\rm C} (\tilde{b}_{i}^\dagger \tilde{b}_{j} + {\rm H.c.})+
        \tilde{V}_1^{\rm C} \tilde{n}_{i}\tilde{n}_{j} \right] \\
        + \sum_{\left\langle\left\langle i,k \right\rangle\right\rangle} \left\{
        \tilde{s}_{1}^{\rm C} \left[ \tilde{b}_{i}^\dagger (\tilde{n}_{j1} + \tilde{n}_{j2})\tilde{b}_{k} +{\rm H.c.} \right] 
        + \tilde{V}_2^{\rm C} \tilde{n}_{i}\tilde{n}_{k} \right\} \\
        + \sum_{\left\langle\left\langle\left\langle i,l \right\rangle\right\rangle\right\rangle} \left[
        \tilde{s}_{2}^{\rm C} \left( \tilde{b}_{i}^\dagger \tilde{n}_{j}\tilde{b}_{l} +{\rm H.c.} \right)
        + \tilde{V}_3^{\rm C} \tilde{n}_{i}\tilde{n}_{l} \right]~.
  \end{split}
  \label{eq:CORE-minimal-hole}
\end{equation}
$\tilde{n}_{i}=\tilde{b}^{\dagger}_{i}\tilde{b}_{i}$ is the occupation number for {\em holes};
$\left\langle i,j \right\rangle$, $\left\langle\left\langle i,k \right\rangle\right\rangle$ and
$\left\langle\left\langle\left\langle i,l \right\rangle\right\rangle\right\rangle$  denote,
respectively, NN, NNN and third-NN sites on the square lattice. The correlated
hopping term with amplitude $\tilde{s}_{1}^{\rm C}$ [$\tilde{s}_{2}^{\rm C}$] is depicted
in Fig.~\ref{fig:leapfrog}(a) [Fig.~\ref{fig:leapfrog}(b)]: a hole hops between two NNN [third-NN]
sites $i$ and $k$ [$l$] only if at least one of their common NN sites $j1$, $j2$ [$j$] is occupied
by a hole.\cite{uncorrelated}

Mean-field results (not shown) for the superfluid density $\rho_{\rm S}$ and the CBS order parameter
$S(\pi,\pi)$ for the minimal model of Eq.~(\ref{eq:CORE-minimal-hole}), with effective couplings
given in Table \ref{tab:revsymb} (for $\Delta=3.3$ and $J/J_\perp=0.29$), semi-quantitatively
reproduce the QMC results (shown in Fig.~\ref{fig:QMC-spin}) for the original spin-dimer model,
Eq.~(\ref{eq:spin-dimer}). Unfortunately, this picture is too simplistic and results from QMC simulations
(not shown) for this minimal model show that the CBS plateau is
destroyed by quantum fluctuations, seemingly invalidating our analysis. However, the QMC results
for $S(\pi,\pi)$ display a rather pronounced peak, indicating that our minimal model is close to a
borderline where the solid phase appears: this is confirmed by the existence of an extended CBS
plateau (concomitantly with a SS phase) in the QMC results obtained by considering
{\em slightly} smaller values for the NN hopping amplitude $\tilde{t}_1^{\rm C}$,\cite{reducing-t1}
suggesting that terms neglected in the full effective model [Eqs.~(\ref{eq:eff_CORE},
\ref{eq:two-body}-\ref{eq:double_hoppings})], although relatively small, play an important role.

A closer examination of the terms in the full effective CORE Hamiltonian
[Eqs.~(\ref{eq:eff_CORE}, \ref{eq:two-body}-\ref{eq:double_hoppings})]
neglected in deriving our minimal model Eq.~(\ref{eq:CORE-minimal-hole})
shows that the NN correlated hoppings with amplitudes $\tilde{s}_3^{\rm C}$
and $\tilde{s}_5^{\rm C}$ [see Eq.~(\ref{eq:corr_hoppings}) and Table \ref{tab:revsymb}]
have exactly the effect of decreasing the holes' (singlets') kinetic energy that may stabilize
the CBS phase. However, the fact that $\tilde{t}_1^{\rm C}$ and $\tilde{s}_3^{\rm C}$, $\tilde{s}_5^{\rm C}$
have opposite signs also implies that their inclusion in Eq.~(\ref{eq:CORE-minimal-hole})
has the undesired effect that the resulting minimal model would suffer from the {\em sign problem}.
In order to circumvent this problem and be able to perform QMC simulations, we
incorporate $\tilde{s}_3^{\rm C}$ and $\tilde{s}_5^{\rm C}$ in an effective way: we
notice that in a perfectly ordered CBS background these extra hoppings effectively reduce
the NN hopping amplitude $\tilde{t}_1^{\rm C}$ to the value we denote $\tilde{t}_1^{\rm min}$
given by (to leading order)
\begin{equation}
  \tilde{t}_1^{\rm min} = \tilde{t}_1^{\rm C} - \left( |\tilde{s}_3^{\rm C}| + |\tilde{s}_5^{\rm C}| \right)~.
  \label{eq:t1_min}
\end{equation}
MF results [Fig.~\ref{fig:MF-CORE}(b)] for the new minimal effective model obtained by the substitution
$\tilde{t}_1^{\rm C} \rightarrow \tilde{t}_1^{\rm min}$ in Eq.~(\ref{eq:CORE-minimal-hole}) suggests
that the dominant physical processes are correctly taken into account, at least close to half-filling,
as we can conclude from the excellent agreement with the results for the full effective CORE model
[Fig.~\ref{fig:MF-CORE}(a)] .\cite{s3s5} Furthermore, the SS region visible in Fig.~\ref{fig:MF-CORE}(b)
is expected to survive quantum fluctuations, for a sizable leapfrog ratio $\Sigma({t}_1^{\rm min}) \approx 0.43$
is obtained for $\Delta=3.3$ and $J/J_\perp=0.29$, something confirmed by our QMC simulations below.

%%%%%%%%%%%%%%%%%%%%%%%%%%%%%%%%%%%%%%%%%%%%%%%%% 
\begin{figure}[!tbp]
  \includegraphics*[width=0.32\textwidth,angle=270]{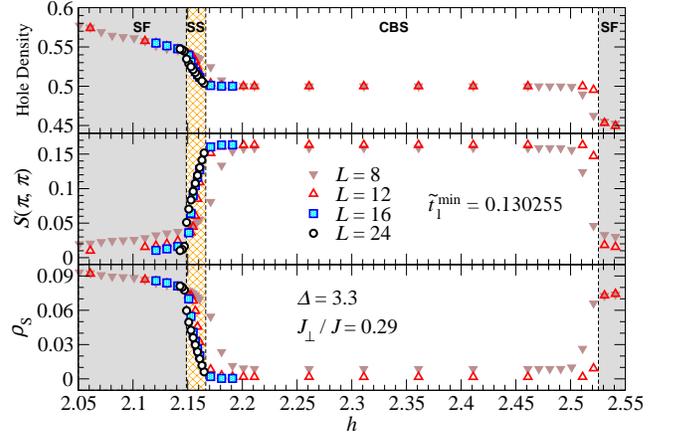}

  \caption{(Color online) $\Delta=3.3$, $J/J_\perp=0.29$. Quantum Monte Carlo results for the 
  minimal effective model obtained from CORE, Eq.~(\ref{eq:CORE-minimal-hole}), considering
  the NN hopping amplitude $\tilde{t}_1^{\rm min}$ from Eq.~(\ref{eq:t1_min}) (values for the
  couplings are given in Table \ref{tab:revsymb}), for lattice sizes $L=8$, $12$, $16$ and $24$.
  Error bars are much smaller than the depicted symbols. (lower panel) Superfluid density, $\rho_{\rm S}$,
  (middle panel) CBS order parameter, $S(\pi,\pi)/N$; (upper panel) singlet density. The temperature is set to
  $T =1/20L< \tilde{t}_1^{\rm min}/2L$ (see main text). Successive phases are labeled as:
  superfluid (${\rm SF}$), supersolid (${\rm SS}$) and checkerboard solid (${\rm CBS}$).
  }
  \label{fig:QMC-minimal}
\end{figure}
%%%%%%%%%%%%%%%%%%%%%%%%%%%%%%%%%%%%%%%%%%%%%%%%% 

\subsection{Quantum Monte Carlo Simulations}
\label{sec:QMC-minimal}

We have performed QMC simulations, using an extended version\cite{schmidt:06} of the
ALPS libraries' implementation\cite{alet:05a,alet:05b} of the Stochastic Series Expansion (SSE)
algorithm.\cite{sandvik:99,sandvik:02} We consider the minimal effective model of Eq.~(\ref{eq:CORE-minimal-hole})
with NN hopping amplitude $\tilde{t}_1^{\rm min}$ given by Eq.~(\ref{eq:t1_min}).
\cite{uncorrelated} We evaluate the superfluid density $\rho_{\rm S}$, obtained in terms
of the winding numbers $w_x$ and $w_y$
\begin{equation}
  \rho_{\rm S}=\frac{1}{2\beta L^2} \langle {w_{\rm x}}^2+{w_{\rm y}}^2\rangle~,
\end{equation}
where $\beta$ is the inverse temperature and $L$ is the system size, and the CBS order parameter
\begin{equation}
  S(\pi,\pi)=\frac{1}{L^2} \langle\sum_{\vec{r}_i,\vec{r}_j}(-1)^{\vec{r}_i-\vec{r}_j}
        \tilde{n}_{\vec{r}_i}\tilde{n}_{\vec{r}_j}\rangle~,
\end{equation}
as a function of the magnetic field $h$. Since we are interested in accessing ground-state properties,
and the main kinetic energy scale in the minimal model Eq.~(\ref{eq:CORE-minimal-hole}) is
$\tilde{t}_1^{\rm min}/J_\perp \approx 0.13$, we set the temperature to $T =1/20L< \tilde{t}_1^{\rm min}/2L$.
It is important to remark that these temperatures are considerably lower than those considered by
Ng and Lee,\cite{ng:06} who assumed that $J/J_\perp =0.29$ was the relevant energy scale, and
this might explain the round shape observed in some of their curves.

QMC results for $\rho_{\rm S}$ and $S(\pi,\pi)$ for the minimal model of Eq.~(\ref{eq:CORE-minimal-hole})
with NN hopping amplitude given by Eq.~(\ref{eq:t1_min}), using the effective couplings appearing
in Table \ref{tab:revsymb} ($\Delta=3.3$ and $J/J_\perp=0.29$), are shown in Fig.~\ref{fig:QMC-minimal}.
The overall agreement with QMC results for the original spin-dimer model Eq.~(\ref{eq:spin-dimer}),
\cite{ng:06,laflorencie:07} shown in Fig.~\ref{fig:QMC-spin}, is good and we can conclude that the minimal
model of Eqs.~(\ref{eq:CORE-minimal-hole},\ref{eq:t1_min}) indeed accurately describes the low-energy physics
of the original model and, more importantly, that the ``leapfrog mechanism" presented in Sec.~\ref{sec:model}
is at least partially responsible for spin-supersolid behavior. 

%However, the extent of the SS phase
%in Fig.~\ref{fig:QMC-minimal} is considerably smaller than for the spin-dimer model.
%Since the {\em sign-problem} precludes us from performing QMC simulations for the full CORE effective
%model, Eqs.~(\ref{eq:eff_CORE}, \ref{eq:two-body}-\ref{eq:double_hoppings}), we can only conjecture and
%some possibilities are: (i) our MF analysis suggests that the effective model obtained from CORE is close
%to a borderline and that terms ignored in obtaining the minimal model, even with small couplings,
%might drastically alter the phase diagram; (ii) the NN correlated hoppings with amplitudes
%$\tilde{s}_3^{\rm C}$, $\tilde{s}_5^{\rm C}$ appear to favor SS\cite{s3s5} and the fact that we
%include only their effects in reducing $\tilde{t}_1^{\rm C}$ [Eq.~(\ref{eq:t1_min})] might be
%responsible for the reduced SS phase in Fig.~\ref{fig:QMC-minimal}; (iii) longer range effective
%interactions and neglected triplet excitations ($\ket{t^{0}}$, $\ket{t^{-1}}$) may be required
%for a better comparison with the results for the original model.

Although the just presented results show that the essential ingredients for spin-supersolidity
have been identified, it is clear that quantitative agreement is not achieved. Specifically, the
extent of the SS phase is considerably smaller in Fig.~\ref{fig:QMC-minimal} than in Fig.~\ref{fig:QMC-spin};
reversely, the CBS phase in the former is about twice as large than in the latter. In trying to understand this
mismatch it is important to keep in mind that supersolidity emerges in this model as the result of a delicate balance
between kinetic and interaction terms. This is evident in the MF analysis discussed in Sec.~\ref{sec:minimal},
which suggests that the effective model obtained from CORE is close to a borderline and that small variations in the
effective couplings can have drastic effects.
%As regard CBS, we indeed showed that if we restrict the Hamiltonian Eq.~\ref{eq:CORE-minimal-hole} to the t-V model terms, the CBS phase simply vanishes whereas from the minimal Hamiltonian Eq.~\ref{eq:CORE-minimal-hole} its extent is twice as large as expected. 
For instance, we have shown that the minimal model
Eq.~(\ref{eq:CORE-minimal-hole}), with effective
couplings shown in Table \ref{tab:revsymb}, does not
display a CBS phase; however, by replacing
$\tilde{t}_1^{\rm C} \rightarrow \tilde{t}_1^{\rm min}$
[Eq.~(\ref{eq:t1_min})] we obtain a CBS phase
twice as large as expected.

 Therefore, and since the
{\em sign-problem} precludes us from performing QMC simulations for the full effective Hamiltonian
[Eqs.~(\ref{eq:eff_CORE}, \ref{eq:two-body}-\ref{eq:double_hoppings})], we conjecture that that terms ignored in
obtaining the minimal model [Eqs.~(\ref{eq:CORE-minimal-hole},\ref{eq:t1_min})], even with small couplings, must be
included in order to better reproduce the results for the original model, shown in Fig.~\ref{fig:QMC-spin}. Additionally, the
NN correlated hoppings with amplitudes $\tilde{s}_3^{\rm C}$ and $\tilde{s}_5^{\rm C}$ appear to favor SS and
the fact that we include only their effects in reducing $\tilde{t}_1^{\rm C}$ [Eq.~(\ref{eq:t1_min})] might be responsible
for the reduced SS phase in Fig.~\ref{fig:QMC-minimal}.\cite{s3s5} Finally, it is not possible to exclude the possibility that longer
range effective interactions and/or neglected triplet excitations ($\ket{t^{0}}$, $\ket{t^{-1}}$) may be required for obtaining
quantitative agreement.

%However, from a quantitative point of view, one has to admit that , first, CBS phase is larger than expected and, second, SS is quite smaller.

%As a general answer, one has
%to realize that the phases and their extensions we try to reproduce here
%are very fragile features because we are actually in a very demanding
%area of parameter space. Both, the CBS and the SS, appear to be weak as
%MF clearly showed.

%As regard CBS, we indeed showed that if we restrict the Hamiltonian Eq.~\ref{eq:CORE-minimal-hole} to the t-V model terms, the CBS phase simply vanishes whereas from the minimal Hamiltonian Eq.~\ref{eq:CORE-minimal-hole} its extent is twice as large as expected. 

%Thus it appears that whereas taking into account correlated hopping terms is sufficient to explain qualitatively the existence of the SS phase, to reach a good quantitative agreement would need to take into account the small couplings we ignore in first approximation.

%Regarding the SS phase, we can also note that  
%the NN correlated hoppings with amplitudes
%$\tilde{s}_3^{\rm C}$, $\tilde{s}_5^{\rm C}$ appear to favor SS and the fact that we
%include only their effects in reducing $\tilde{t}_1^{\rm C}$ [Eq.~(\ref{eq:t1_min})] might be
%responsible for the reduced SS phase in Fig.~\ref{fig:QMC-minimal} as confirmed by MF calculations.\cite{s3s5} We can of course not exclude that longer range effective
%interactions and neglected triplet excitations ($\ket{t^{0}}$, $\ket{t^{-1}}$) may be required
%for a better comparison with the results for the original model.

%%%%%%%%%%%%%%%%%%%%%%%%%%%%%%%%%%%%%%%%%%%%%%%%% 
\begin{figure}[!tbp]
  \includegraphics*[width=0.45\textwidth]{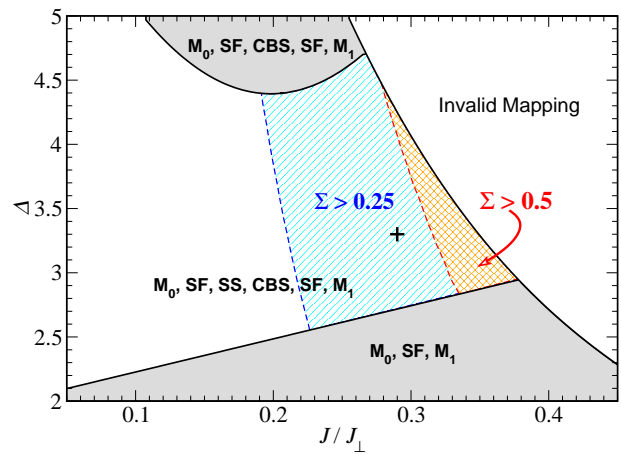}

  \caption{(Color online)
  Successive phases stabilized for increasing magnetic field $h$ in the parameter space of
  the spin-dimer model of Eq.~(\ref{eq:spin-dimer}), as obtained from a mean-field analysis of the full
  CORE Hamiltonian, Eqs.~(\ref{eq:eff_CORE}, \ref{eq:two-body}-\ref{eq:double_hoppings}).
  Grey shaded areas correspond to parameters for which no supersolid phase is found and in the
  region marked as ``invalid mapping" the CORE expansion is invalid (see Sec.~\ref{sec:CORE_procedure}).
  In the remaining area a supersolid phase is obtained within the mean-field approach. The colorful regions indicate
  parameters for which the singlet ``leapfrog ratio"  [Eq.~(\ref{eq:leapfrog_ratio})] is larger
  than the threshold value $\Sigma({\tilde{t}}_1^{\rm min}) > c/4$ (with $c \in [1,2]$), as required for SS phases to appear.
  The cross highlights parameters $\Delta=3.3$ and $J/J_\perp=0.29$ from Refs.~\onlinecite{ng:06},
  \onlinecite{laflorencie:07}. Phases are labeled as: Mott insulator (${\rm M}_0$, empty, and
  ${\rm M}_1$, full), superfluid (${\rm SF}$), supersolid (${\rm SS}$) and checkerboard solid (${\rm CBS}$).
  }
  \label{fig:MF_phase_diagram}
\end{figure}
%%%%%%%%%%%%%%%%%%%%%%%%%%%%%%%%%%%%%%%%%%%%%%%%%

\subsection{Extent of the Supersolid Phase}
\label{sec:extent}

We have extended our MF analysis to the full effective model Eqs.~(\ref{eq:eff_CORE},
\ref{eq:two-body}-\ref{eq:double_hoppings}) by varying the parameters $\Delta$ and $J/J_\perp$,
as a function of the magnetic field $h$. Results are shown in Fig.~\ref{fig:MF_phase_diagram}.
Grey shaded areas represent values of $\Delta$ and $J/J_\perp$ for which no SS phase is stabilized within
MF for all values of $h$ and only a superfluid and/or CBS phases are obtained. However, SS does appear
over an extended region in the parameters space within MF. Since it is well known
that MF tends to overestimate supersolidity, we have also analyzed the ``leapfrog ratio" $\Sigma({t}_1^{\rm min})$
[see Eq.~(\ref{eq:leapfrog_ratio})] throughout the parameters space. As discussed in Sec.~\ref{sec:model},
the condition $\Sigma({t}_1^{\rm min}) > c/4$, with $c \in [1,2]$, must be satisfied for preventing phase separation
and stabilizing a SS. Regions for which this condition is fulfilled are indicated in Fig.~\ref{fig:MF_phase_diagram}:
we can see that the region where the SS phase is likely to occur is much smaller than expected from
the MF analysis and, in particular, no SS is expected within the perturbative limit.

\section{Conclusions and Outlook}

Summarizing, we have obtained effective models describing the low-energy physics of a spin-dimer
model [Eq.~(\ref{eq:spin-dimer})] known to exhibit spin-supersolid behavior\cite{ng:06,laflorencie:07}
with the help of perturbative expansions and of the contractor renormalization (CORE) algorithm.
While the perturbative analysis, relying on the assumption of a disordered ground state with gapped
excitations at zero-field, does not reproduce the extended
supersolid phase observed in the original model
(Fig.\ref{fig:QMC-spin}), CORE does not assume any particular
ordering in the system and is shown to reproduce the main features obtained from more computationally
demanding approaches, even when a simple mean-field procedure is applied to the obtained effective
model.

Furthermore, 
%as an important byproduct, CORE allows us to gauge the relative importance of terms in
%the effective Hamiltonian and therefore to
we identify the mechanism at play behind spin-supersolidity and we show that the spin-supersolid phase exhibited by the $S=1/2$ spin-dimer model
of Eq.~(\ref{eq:spin-dimer}) can be simply understood in terms of the ``leapfrog mechanism" illustrated
in Fig.~\ref{fig:leapfrog}. Basically, a sizable amplitude for correlated hoppings allow extra holes (singlets)
to delocalize on the other sublattice of a checkerboard solid, preventing phase separation and
leading to supersolid behavior.

More generally speaking, we are able to describe the
physics behind complex phenomena in a simple
way by deriving effective models with only a few terms
and rather local couplings. The essential physical
ingredients can be identified even in a low-order
perturbative analysis, although more sophisticated
approaches, such as PCUTs and CORE, may be required in
obtaining the effective couplings. We highlight that
both PCUTs and CORE are immune to the {\em sign problem}
and can therefore be applied to frustrated and fermionic
systems, something which opens interesting research
possibilities.

%More generally speaking, even in non perturbative situations,
%we are able to describe the physics in a quite
%simple way i.e. using a simple model with rather local
%terms. But, even if we have the same parameters than in
%the low order perturbative regime, we need tools to find
%quantitatively the renormalization of the local couplings.
%This can in general be done by PCUT or CORE (both immune to the {\em sign problem} which plagues QMC simulations). In the
%present case we have to use CORE since it allows to deduce
%a low-energy effective model even in the presence
%of quantum phase transitions. We also want to underscore
%the efficiency of working with effective models to
%get more insight in the physics of a given harder model.
%This could even be more relevant in cases where there are
%no other methods to deal with the original spin model.

%Finally, we stress that CORE is immune to the {\em sign problem} which plagues QMC simulations,
%a feature particularly advantageous if one is interested in spin-supersolid behavior: it is known that
%frustration reduces the kinetic energy of emergent bosons in dimerized magnets, eventually stabilizing
%solid phases\cite{Mila:1998,abendschein:07} and, possibly, spin-supersolidity. It is therefore interesting
%to employ CORE in trying to identify realistic candidates for spin-supersolidity, in which the artificially large
%Ising anisotropy $\Delta$ of the model considered here [Eq.~(\ref{eq:spin-dimer})] is replaced by frustration.

\begin{acknowledgments}
We acknowledge the financial support of the Swiss National Fund and of MaNEP. 

A.~F.~Albuquerque is indebted to A.~Abendschein for sharing numerical results and for fruitful discussions
  and acknowledges financial support from CNPq (Brazil), NIDECO (Switzerland) and ARC (Australia).

K.~P.~Schmidt acknowledges ESF and EuroHorcs for funding through his EURYI.

N. L. acknowledges the Laboratoire de physique th\'eorique (Toulouse)
for hospitality.

F.~M. acknowledges the r\'egion Midi-Pyr\'en\'ees through its Chaire d'excellence Pierre de Fermat program for financial support and thanks the Laboratoire de physique th\'eorique (Toulouse) for hospitality.
\end{acknowledgments}

\appendix
 
 %%%%%%%%%%%%%%%%%%%%%%%%%%%%%%%%%%%%%%%%%%%%%%%%% 
 
\section{PCUTs sixth-order effective couplings}
\label{sec:appendix-PCUT}

The effective couplings obtained from the PCUTs analysis discussed in Sec.~\ref{sec:PCUTs} are
\begin{equation}
 \begin{split}
t_1^{\rm P6}= \dfrac{1}{2} J - \dfrac{5}{32} J^3 - \dfrac{3}{32} J^3 \Delta^2 - \dfrac{13}{32} J^4 \Delta - \dfrac{1}{16} J^4 \Delta^3 \\
-\dfrac{265}{512} J^5 - \dfrac{61}{512} J^5 \Delta^2  \dfrac{57}{256} J^5 \Delta^4 -\dfrac{393}{1024} J^6 \Delta \\
+   \dfrac{193}{1024} J^6 \Delta^3 - \dfrac{15}{64} J^6 \Delta^5 \nonumber
 \end{split}
\end{equation}
\begin{equation}
 \begin{split}
V_1^{\rm P6}= \dfrac{1}{2} J \Delta- \dfrac{1}{4} J^2 - \dfrac{1}{8} J^2 \Delta^2 -  \dfrac{3}{16} J^3 \Delta  - \dfrac{95}{128} J^4  \\
  + \dfrac{79}{128} J^4 \Delta^2-\dfrac{51}{128} J^4 \Delta^4 -  \dfrac{395}{1024} J^5 \Delta \\
  +\dfrac{191}{1024} J^5 \Delta^3  - \dfrac{14937}{8192} J^6 + \dfrac{11545}{16384} J^6 \Delta^2 \\
  + \dfrac{32641}{16384} J^6 \Delta^4 - \dfrac{419}{256} J^6 \Delta^6  \nonumber
 \end{split}
\end{equation}
\begin{equation}
 \begin{split}
t_2^{\rm P6}= -\dfrac{1}{4} J^2 - \dfrac{1}{4} J^3 \Delta + \dfrac{1}{16} J^4 + \dfrac{1}{32} J^4 \Delta^2 + 
  \dfrac{1}{2} J^5 \Delta   \\
  + \dfrac{3}{32} J^5 \Delta^3 + \dfrac{63}{128} J^6 + \dfrac{45}{256} J^6 \Delta^2 + 
  \dfrac{171}{512} J^6 \Delta^4 \nonumber
 \end{split}
\end{equation}
\begin{equation}
 \begin{split}
  t'^{\rm P6}_2= \dfrac{1}{8} J^2 +\dfrac{5}{32} J^3 \Delta - \dfrac{35}{256} J^4 - \dfrac{3}{128} J^4 \Delta^2 -\dfrac{1089}{2048} J^5 \Delta \\
  - \dfrac{77}{1024} J^5 \Delta^3 - \dfrac{3025}{4096} J^6 - \dfrac{3623}{16384} J^6 \Delta^2 - \dfrac{5737}{16384} J^6 \Delta^4  \nonumber
 \end{split}
\end{equation}
%
%%%%%%%%%%%%%%%%%%%%%%%%%%%%%%%%%%%%%%%%%%%%%%%%% 

%%%%%%%%%%%%%%%%%%%%%%%%%%%%%%%%%%%%%%%%%%%%%%%%% 
\section{Effective couplings from CORE}
\label{sec:appendix-CORE}

The explicit expressions for each term in the effective Hamiltonian obtained from CORE, Eq.~(\ref{eq:eff_CORE}),
are given here (in the expressions below $\tilde{n}_{i}=\tilde{b}^{\dagger}_{i}\tilde{b}_{i}$ is the occupation number
for {\em holes} and $\hat{x}, \hat{y}$ are unity vectors for the square lattice; constant terms arising from applying a particle-hole
transformation to the bare CORE effective Hamiltonian are ignored) . $\tilde{{\mathcal V}}$ comprises two-body interactions
\begin{equation}
  \begin{split}
    \tilde{{\mathcal V}}_{i} &=     \tilde{V}_1^{\rm C} ( \tilde{n}_{i}\tilde{n}_{i + \hat{x} } + \tilde{n}_{i}\tilde{n}_{i + \hat{y} } )\\
    &+ \tilde{V}_2^{\rm C} ( \tilde{n}_{i}\tilde{n}_{i + \hat{x} + \hat{y} } + \tilde{n}_{i}\tilde{n}_{i + \hat{x} - \hat{y} } )\\
    &+ \tilde{V}_3^{\rm C} ( \tilde{n}_{i}\tilde{n}_{i + 2 \hat{x} } + \tilde{n}_{i}\tilde{n}_{i + 2 \hat{y} } )~,
  \end{split}
  \label{eq:two-body}
\end{equation}
and $\tilde{{\mathcal W}}$ three- and four-body interactions
\begin{equation}
  \begin{split}
    &\tilde{\mathcal{W}}_{i} =  \tilde{W}_1^{\rm C} ( \tilde{n}_{i}\tilde{n}_{i + \hat{x}}\tilde{n}_{i + 2\hat{x}} + \tilde{n}_{i}\tilde{n}_{i + \hat{y}}\tilde{n}_{i + 2\hat{y}} )\\
    &+ \tilde{W}_2^{\rm C} [ \tilde{n}_{i}\tilde{n}_{i + \hat{x}} ( \tilde{n}_{i + \hat{y}} + \tilde{n}_{i + \hat{x} + \hat{y}} +\tilde{n}_{i - \hat{y}} + \tilde{n}_{i + \hat{x} - \hat{y}} ) ]\\
    &+ \tilde{W}_3^{\rm C} ( \tilde{n}_{i}\tilde{n}_{i + \hat{x}}\tilde{n}_{i + \hat{y}}\tilde{n}_{i + \hat{x} + \hat{y}} ).
  \end{split}
  \label{eq:multi-body}
\end{equation}
The effective single-boson hopping terms in Eq.~(\ref{eq:eff_CORE}) are
\begin{equation}
  \begin{split}
    \tilde{{\mathcal T}}_{i}& =      \tilde{t}_1^{\rm C} (\tilde{b}^{\dagger}_{i}\tilde{b}_{i + \hat{x} } + \tilde{b}^{\dagger}_{i}\tilde{b}_{i + \hat{y} } + {\rm H.c.} )\\
    &+ \tilde{t}_2^{\rm C} (\tilde{b}^{\dagger}_{i}\tilde{b}_{i + \hat{x} + \hat{y} } + \tilde{b}^{\dagger}_{i}\tilde{b}_{i + \hat{x} - \hat{y} } + {\rm H.c.} )~.
  \end{split}
  \label{eq:single_hoppings}
\end{equation}
Correlated hopping terms are
\begin{equation}
  \begin{split}
    \tilde{{\mathcal S}}_{i} =&\\ 
    \tilde{s}_1^{\rm C} [&\tilde{b}^{\dagger}_{i}(\tilde{n}_{i + \hat{x} } + \tilde{n}_{i + \hat{y} })\tilde{b}_{i + \hat{x} + \hat{y} }\\
    +& \tilde{b}^{\dagger}_{i}(\tilde{n}_{i + \hat{x} } + \tilde{n}_{i - \hat{y} })\tilde{b}_{i + \hat{x} - \hat{y} } + {\rm H.c.} ]\\
    + \tilde{s}_2^{\rm C} [&\tilde{b}^{\dagger}_{i}\tilde{n}_{i + \hat{x} }\tilde{b}_{i + 2\hat{x} } + \tilde{b}^{\dagger}_{i}\tilde{n}_{i + \hat{y} }\tilde{b}_{i + 2 \hat{y} } + {\rm H.c.} ]\\
    + \tilde{s}_3^{\rm C} [&\tilde{b}^{\dagger}_{i}\tilde{b}_{i + \hat{x} } (\tilde{n}_{i + \hat{y} } + \tilde{n}_{i + \hat{x} + \hat{y} } + \tilde{n}_{i - \hat{y} } + \tilde{n}_{i + \hat{x} - \hat{y} } )\\
    +& \tilde{b}^{\dagger}_{i}\tilde{b}_{i + \hat{y} } (\tilde{n}_{i + \hat{x} } + \tilde{n}_{i + \hat{x} + \hat{y} } + \tilde{n}_{i - \hat{x} } + \tilde{n}_{i - \hat{x} + \hat{y} } ) + {\rm H.c.} ]\\
    + \tilde{s}_4^{\rm C} [&\tilde{b}^{\dagger}_{i}\tilde{b}_{i + \hat{x} } (\tilde{n}_{i + \hat{y} }\tilde{n}_{i + \hat{x} + \hat{y} } + \tilde{n}_{i - \hat{y} }\tilde{n}_{i + \hat{x} - \hat{y} } )\\
    +& \tilde{b}^{\dagger}_{i}\tilde{b}_{i + \hat{y} } (\tilde{n}_{i + \hat{x} }\tilde{n}_{i + \hat{x} + \hat{y} } + \tilde{n}_{i - \hat{x} }\tilde{n}_{i - \hat{x} + \hat{y} } ) + {\rm H.c.} ]\\
    + \tilde{s}_5^{\rm C} [&\tilde{b}^{\dagger}_{i}\tilde{b}_{i + \hat{x} } (\tilde{n}_{i - \hat{x} } + \tilde{n}_{i + 2 \hat{x} })\\
    +&\tilde{b}^{\dagger}_{i}\tilde{b}_{i + \hat{y} } (\tilde{n}_{i - \hat{y} } + \tilde{n}_{i +2 \hat{y} }) + {\rm H.c.} ]\\
    + \tilde{s}_6^{\rm C} (&\tilde{b}^{\dagger}_{i}\tilde{n}_{i + \hat{x} }\tilde{n}_{i + \hat{y} }\tilde{b}_{i + \hat{x} + \hat{y} }
    + \tilde{b}^{\dagger}_{i}\tilde{n}_{i + \hat{x} }\tilde{n}_{i - \hat{y} }\tilde{b}_{i + \hat{x} - \hat{y} } + {\rm H.c.} )~,
  \end{split}
  \label{eq:corr_hoppings}
\end{equation}
and, finally, hoppings simultaneously involving two-bosons
\begin{equation}
  \begin{split}
    \tilde{{\mathcal R}}_{i} =      \tilde{r}_1^{\rm C} (&\tilde{b}^{\dagger}_{i}\tilde{b}^{\dagger}_{i + \hat{x}}\tilde{b}_{i + \hat{y}}\tilde{b}_{i + \hat{x} +  \hat{y}}  \\
    +  &\tilde{b}^{\dagger}_{i}\tilde{b}^{\dagger}_{i + \hat{y}}\tilde{b}_{i + \hat{x}}\tilde{b}_{i + \hat{x} +  \hat{y}} + {\rm H.c.} )\\
    +    \tilde{r}_2^{\rm C} (&\tilde{b}^{\dagger}_{i}\tilde{b}_{i + \hat{x}}\tilde{b}_{i + \hat{y}}\tilde{b}^{\dagger}_{i + \hat{x} + \hat{y}}\\
    + &\tilde{b}^{\dagger}_{i}\tilde{b}_{i + \hat{x}}\tilde{b}_{i - \hat{y}}\tilde{b}^{\dagger}_{i + \hat{x} - \hat{y}} + {\rm H.c.} )~.
  \end{split}
  \label{eq:double_hoppings}
\end{equation}

%%%%%%%%%%%%%%%%%%%%%%%%%%%%%%%%%%%%%%%%%%%%%%%%% 
\section{Mean-field Procedure}
\label{sec:appendix-MF}

Following the Matsubara-Matsuda semiclassical approach,\cite{matsubara:56} we write the hard-core boson effective models in terms of $S = 1/2$ pseudo-spin variables. We start by replacing the commutation relations
for bosons on the same site $i$,

\begin{equation}
[b_i,b_i]=[b^\dagger_i,b^\dagger_i]=0\quad \textrm{and} \quad [b_i,b^\dagger_i]=1
\end{equation}

by the fermionic anticommutation relations

\begin{equation}
\{b_i,b_i\}=\{b^\dagger_i,b^\dagger_i\}=0\quad \textrm{and} \quad\{b_i,b^\dagger_i\}=1
\end{equation}

\noindent while retaining the canonical bosonic commutators for
operators on different sites $i$, $j$. This leads to an algebra formally equivalent to that of a spin $1/2$.

We then neglect quantum fluctuations by replacing the pseudo-spin operators by their mean value, obtaining
a Hamiltonian in terms of classical spins variables $\mathbf{S}=(\cos\phi \sin\theta, \sin\phi \sin\theta, \cos\theta)$ which reads 
\begin{equation}
  \begin{split}
        \mathcal{H}_{\rm MF} = h_{\rm eff} \sum_{i} S_i^z
        + \sum_{\left\langle i,j \right\rangle} \left[ \tilde{J}_{ij}^{z}  S_{i}^{z}S_{j}^{z}
        + \tilde{J}_{ij}^{\perp}(S_i^xS_{j}^x+S_i^yS_{j}^y) \right] \\
        +\sum_{\left\langle i,j,k \right\rangle} \left[ \tilde{K}^{\perp}_{ijk} S_{i}^z(S_j^xS_{k}^x+S_j^yS_{k}^y ) 
        + \tilde{K}^{z}_{ijk} S_{i}^zS_j^zS_{k}^z\right] \\
        +\sum_{\left\langle i,j,k,l \right\rangle} \left[ \tilde{L}^{\perp}_{ijkl} S_{i}^zS_{j}^z(S_k^xS_{l}^x+S_k^yS_{l}^y ) 
        + \tilde{L}^{z}_{ijkl} S_{i}^zS_j^zS_{k}S_{l}^z\right] \\
        +\sum_{\left\langle i,j,k,l \right\rangle} \left[ \tilde{M}^{\perp}_{ijkl} (S_{i}^{+}S_{j}^{-}S_k^{+}S_{l}^{-}+{\rm H.c.}) \right] ~.
  \end{split}
\end{equation}
The parameters $h_{\rm eff}$ (one-body), $\tilde{J}$ (two-body), $\tilde{K}$ (three-body),
$\tilde{L}$ (four-body) and $\tilde{M}$ (double exchange) are defined in terms of the couplings
in the effective bosonic Hamiltonian. The superscript $z$ ($\perp $) accounts for interactions (hoppings)
between sites coupled as in the bosonic Hamiltonian.

In accounting for the different phases of the Hamiltonian Eq.~(\ref{eq:spin-dimer}) it suffices to consider a site-factorized
wave-function $\ket{\psi}=\prod_{i}\ket{\psi_i}$ assuming two-sublattice long-range order ($i=A, B$). The variational
parameters ($\phi_A$, $\phi_B$, $\theta_A$ and $\theta_B$) are determined by minimizing the ground-state
 energy per site
%
%\begin{equation}
%  \bra{\psi}\mathcal{H}\ket{\psi}=\dfrac{1}{2}(\bra{\psi_A}\mathcal{H}\ket{\psi_A}+\bra{\psi_B}\mathcal{H}\ket{\psi_B})~.
%\end{equation}
within this subspace. The condensate density corresponds in a MF approach to the magnetization in the $xy$ plane
\begin{equation}
 \begin{split}
  \rho_0= \frac{1}{8}(\sin^2\theta_A
  +\sin^2\theta_B)~,
  \label{eq:MF_rhoS}
 \end{split}
\end{equation}
and the CBS structure factor is 
\begin{equation}
  S(\pi,\pi) = (\cos \theta_A-\cos\theta_B)^2/4~.
  \label{eq:MF_CBS}
\end{equation}

 In terms of the density of singlets in the sublattice $A$,
given by

\begin{equation}
n_A=\frac{1+\cos\theta_A}{2}
\end{equation}
 
\noindent with a similar definition for the sublattice $B$ ($n_B$),
the ground state energy per site $E_0$ (up to a constant)
for the minimal model of Eqs.~(\ref{eq:CORE-minimal-hole}-\ref{eq:t1_min}).

%

%up to a constant, the minimal Hamiltonian $\ref{eq:CORE-minimal-hole}$ (including $s_3$ and $s_5$ contributions for the sake of completeness) (cf. Fig.~\ref{fig:MF-CORE}) leads to the energy $E_0$
 
\begin{align}\nonumber
&E_0=\\
&2\tilde{V}_1^{\rm C}n_An_B\label{eq:c6}\\
&+(\tilde{V}_2^{\rm C}+\tilde{V}_3^{\rm C})({n_A}^2+{n_B}^2)\label{eq:c7}\\
&+4 \tilde{t}_1^{\rm C}\sqrt{n_A(1-n_A)}\sqrt{n_B(1-n_B)}\cos(\phi_A-\phi_B)\label{eq:c8}\\
&+2(\tilde{s}_2^{\rm C}+2\tilde{s}_1^{\rm C})n_An_B(2-n_B-n_A)\label{eq:c9}\\
%&+4(s_3+2s_5)\sqrt{n_An_B(1-n_B)(1-n_A)}\cdot\\
%&\hspace*{3.44cm}(n_A+n_B)\cos(\phi_A-\phi_B)\label{eq:c10}\\
&+(h-\mu)\frac{n_A+n_B}{2}
\end{align}

We can therefore deduce the following trends:

  %%%%%%%%%%%%%%%%%%%%%%%%%%%%%%%%%%%%%%%%%%%%%%%%%%%%%%%%%%%%%%%%

\begin{figure*}[!tbp]
\includegraphics*[width=\columnwidth]{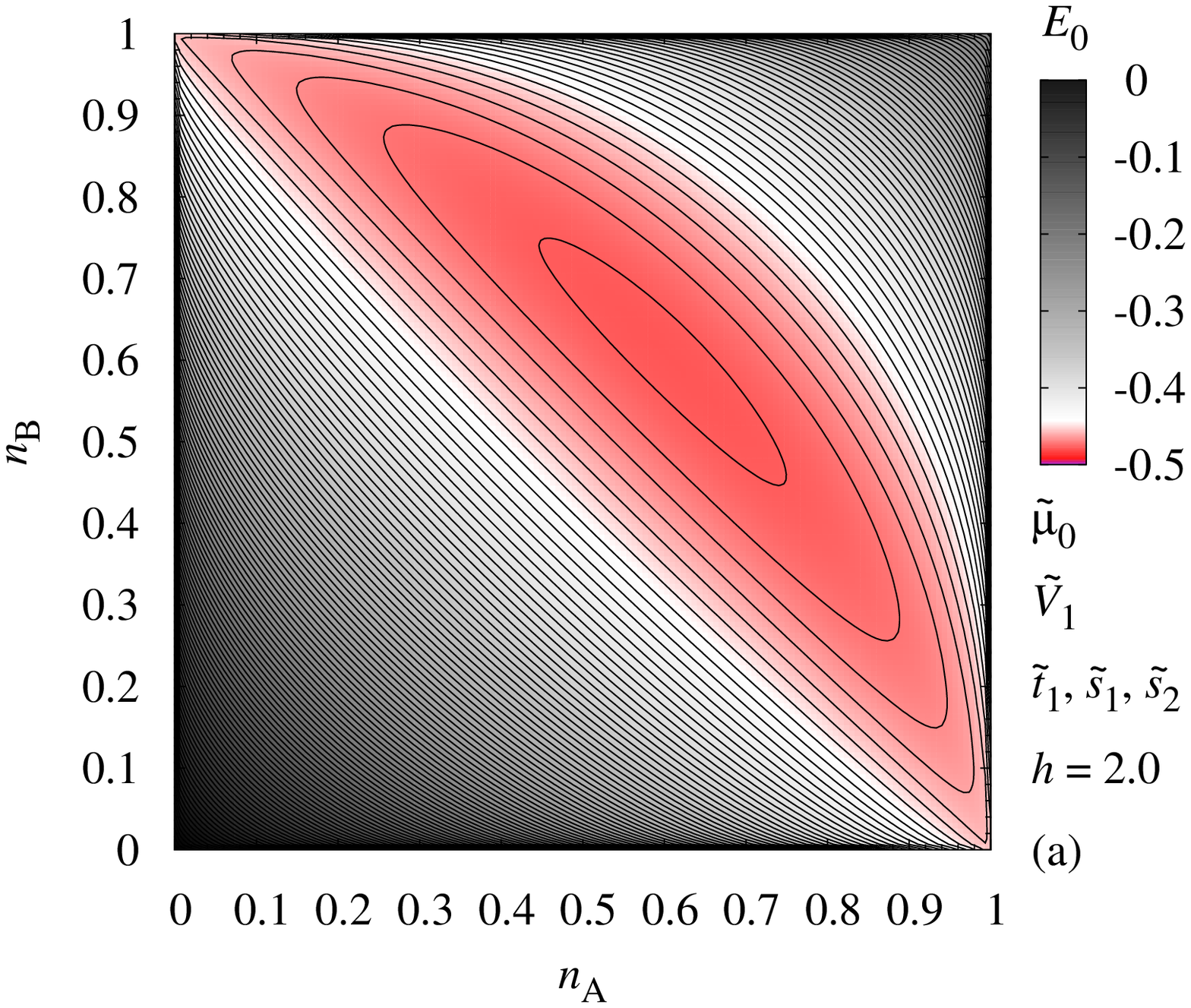}
\includegraphics*[width=\columnwidth]{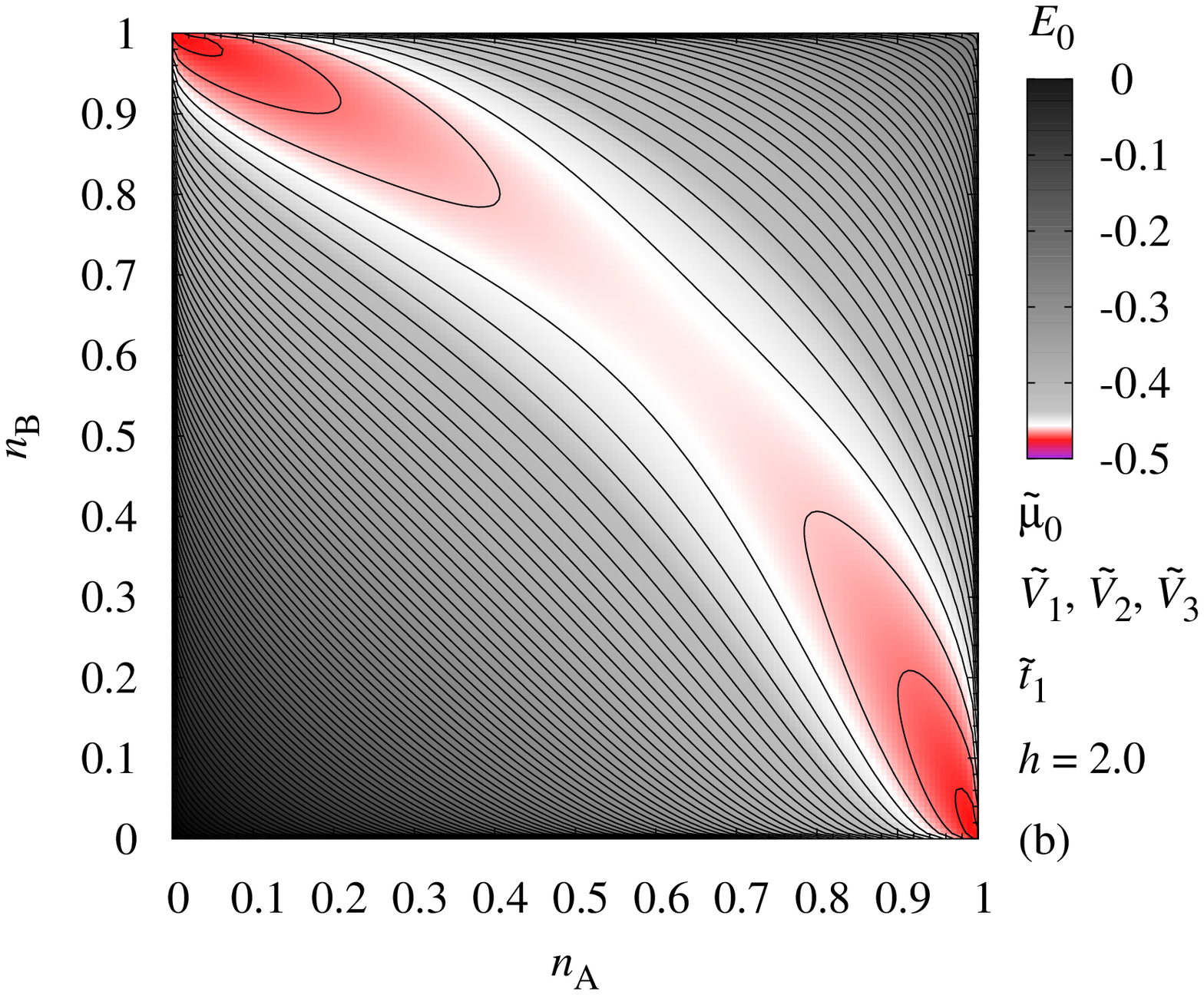}
\includegraphics*[width=\columnwidth]{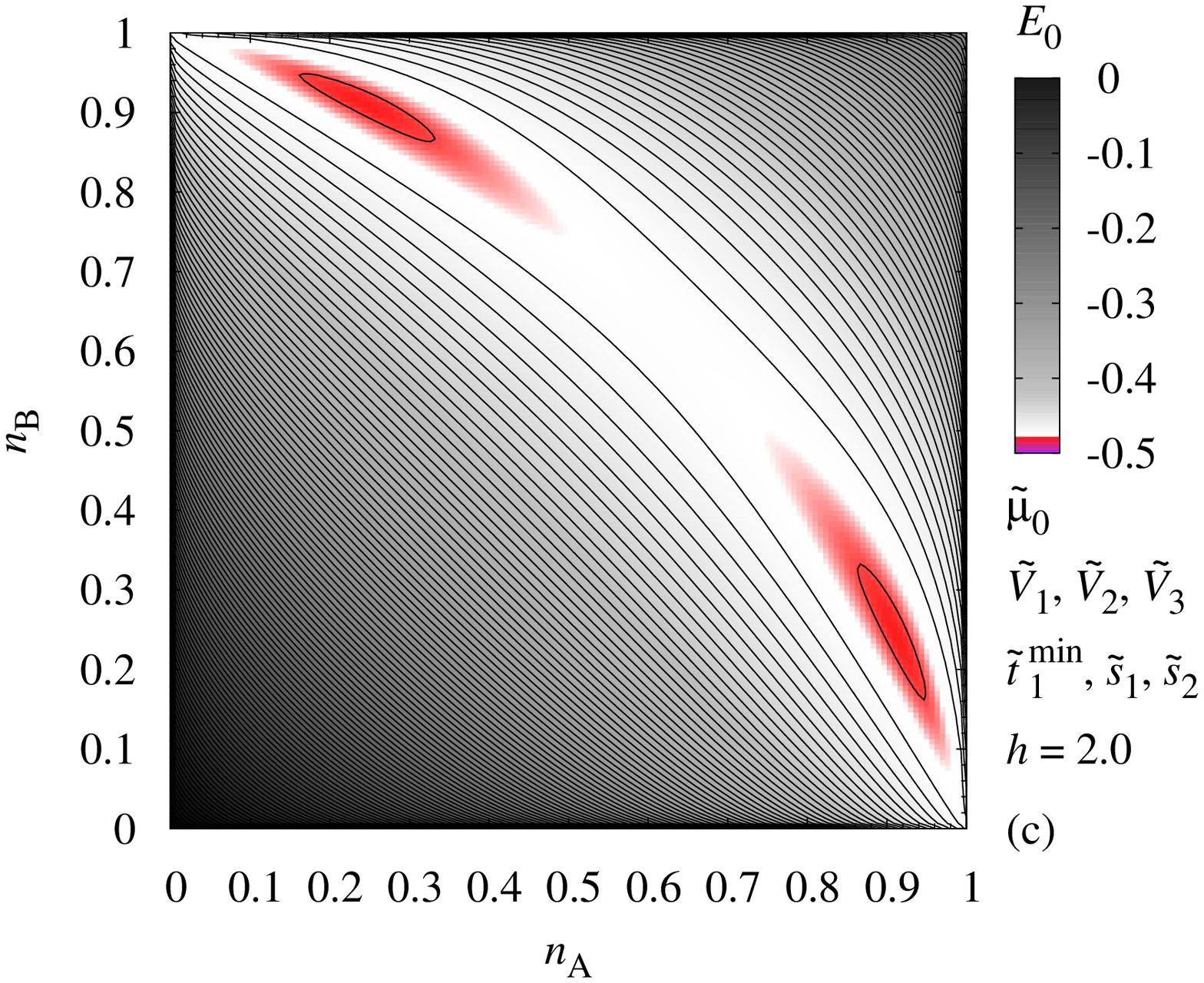}

\caption{(Color online) Mean-field ground-state energy per site, $E_0$,
for $\Delta=3.3$, $J/J_\perp=0.29$ and magnetic
field $h=2$, using the effective couplings shown in
Table \ref{tab:revsymb}, for models comprising the following terms
[see (\ref{eq:CORE-minimal-hole})]: (a) $\tilde{\mu}^{\rm C}$, 
$\tilde{V}_1^{\rm C}$, $\tilde{t}_1^{\rm C}$, $\tilde{s}_1^{\rm C}$ and
$\tilde{s}_2^{\rm C}$, leading to a superfluid phase. (b) $\tilde{\mu}^{\rm C}$,
$\tilde{V}_1^{\rm C}$, $\tilde{V}_2^{\rm C}$, $\tilde{V}_3^{\rm C}$
and $\tilde{t}_1^{\rm C}$, leading to a CBS phase. (c)
The minimal model from Eqs.~(\ref{eq:CORE-minimal-hole},
\ref{eq:t1_min}), for which a spin-supersolid phase is
obtained [cf. Fig.~\ref{fig:MF-CORE}(b)].
%(a) MF energy corresponding to the sum of Eq.~(\ref{eq:c6}) and Eq.~(\ref{eq:c7}) vs densities $n_A$ and $n_B$. The minimum of the energy is not on the diagonal : this terms contribute to a symmetry breaking between $A$ and $B$ sites. (b) MF energy due to hopping term Eq.~(\ref{eq:c8}). The minimum of the energy corresponds to  $n_A=n_B$. (c) MF energy corresponding to the sum of Eq.~(\ref{eq:c9}) and Eq.~(\ref{eq:c10}). This term also breaks the symmetry between $A$ and $B$ sites. (d)  Total MF energy  vs densities $n_A$ and $n_B$. The minimum of the energy corresponds to  $n_A=0$ and $n_B=0.714$.
}
\label{fig:ground-state}
\end{figure*}

%%%%%%%%%%%%%%%%%%%%%%%%%%%%%%%%%%%%%%%%%%%%%%%%%%%%%%%%%%%%%%%%

 \begin{itemize}
\item Due to the sign of $\tilde{V}_1^{\rm C}$, $\tilde{V}_2^{\rm C}$ and $\tilde{V}_3^{\rm C}$, the terms of Eq.~(\ref{eq:c6}) and Eq.~(\ref{eq:c7}) favor the CBS because in order to minimize them, one must break $A-B$ symmetry (cf. Fig.~\ref{fig:ground-state} (b)).

%%%%%%%%%%%%%%%%%%%%%%%%%%%%%%%%%%%%%%%%%%%%%%%%%% 
%\begin{figure}[!tbp]
%  \includegraphics*[width=0.45\textwidth]{V1_V2_V3.eps}
%  \caption{(Color online)
%  MF energy  corresponding to the sum of Eq.~(\ref{eq:c6}) and Eq.~(\ref{eq:c7}) vs densities $n_A$ and $n_B$. The minimum of the energy is not on the diagonal : this terms contribute to a symmetry breaking.}
%  \label{fig:Ep}
%\end{figure}
%%%%%%%%%%%%%%%%%%%%%%%%%%%%%%%%%%%%%%%%%%%%%%%%%%

\item On the contrary, the kinetic term  Eq.~(\ref{eq:c8}) (cf. Fig.~\ref{fig:ground-state} (a)) favors the SF phase : it is indeed minimal for $n_A=n_B$ and $\phi_A-\phi_B=\pi$. There is then no symmetry breaking between $A$ and $B$ sublattices and the latter relation introduces an order in the $xy$ plane confirming the presence of a SS phase.

%%%%%%%%%%%%%%%%%%%%%%%%%%%%%%%%%%%%%%%%%%%%%%%%%% 
%\begin{figure}[!tbp]
% \includegraphics*[width=0.45\textwidth]{t1.eps}

%  \caption{(Color online)
%  MF energy vs densities $n_A$ and $n_B$ corresponding to hopping term Eq.~(\ref{eq:c8})  vs densities $n_A$ and $n_B$. The minimum of the energy corresponds to  $n_A=n_B$.}
%  \label{fig:Ec}
%\end{figure}
%%%%%%%%%%%%%%%%%%%%%%%%%%%%%%%%%%%%%%%%%%%%%%%%%%

\item The last  term Eq.~(\ref{eq:c9}) is more subtle. In the case where $\tilde{s}_1^{\rm C}$ and $\tilde{s}_2^{\rm C}$ are negative as presently, the contribution of Eq.~(\ref{eq:c9}) does not break the translational symmetry and therefore only favors the SF phase. But it is yet sufficient to induce a SS as shown on Fig.~\ref{fig:ground-state} (c). This figure shows that the minimization of the total MF energy indeed leads to translational symmetry breaking ($n_A\neq n_B$) but as the highest density of both is not equal to one, we do not obtain a CBS phase but the SS one.  What is not shown is that minimizing $E_0$  also leads to $\phi_A-\phi_B=\pi$ or in other words to an order in the $xy$ plane which is the semiclassical equivalent of the condensate density. 

\end{itemize}

For the sake of completeness, let's mention that if we would have considered the contributions of $\tilde{s}_3^{\rm C}$ and $\tilde{s}_5^{\rm C}$  indirectly inserted in the minimal Hamiltonian due to the sign problem, they would have led to an MF energy 

\begin{align}\nonumber
%&E_0=\\
%&2\tilde{V}_1^{\rm C}n_An_B\label{eq:c6}\\
%&+(\tilde{V}_2^{\rm C}+\tilde{V}_3^{\rm C})({n_A}^2+{n_B}^2)\label{eq:c7}\\
%&+4 \tilde{t}_1^{\rm C}\sqrt{n_A(1-n_A)}\sqrt{n_B(1-n_B)}\cos(\phi_A-\phi_B)\label{eq:c8}\\
%&+2(\tilde{s}_2^{\rm C}+2\tilde{s}_1^{\rm C})n_An_B(2-n_B-n_A)\label{eq:c9}\\
&+4(\tilde{s}_3^{\rm C}+2\tilde{s}_5^{\rm C})\sqrt{n_An_B(1-n_B)(1-n_A)}\cdot\\\nonumber
&\hspace*{3.44cm}(n_A+n_B)\cos(\phi_A-\phi_B)\label{eq:c10}\\
%&+(h-\mu)\frac{n_A+n_B}{2}
\end{align}

In order to minimize it, $\phi_A-\phi_B$ must be equal either to 0 or to $\pi$ which introduces an order in the $xy$ plane.  Moreover, if $\phi_A-\phi_B=\pi$ (as actually imposed by Eq.~\ref{eq:c8}), it also leads to a breaking of the $A-B$ symmetry as shown by Fig.~\ref{fig:ground-state} (c). At the MF level, this correlated hopping term alone favors both symmetry breakings contrary to Eq.~(\ref{eq:c9}).

\bibliographystyle{apsrev}

\bibliography{References,Comments}

\begin{thebibliography}{51}
\bibitem[{sec()}]{second-order}
\bibinfo{note}{To second-order in $J/J_\perp$, the processes appearing in
  Fig.~\ref{fig:leapfrog}(a) and (b) have the same amplitude. This is not the
  case for the effective Hamiltonian obtained from CORE, see
  Table~\ref{tab:revsymb}.}
\end{thebibliography}

\end{document}